\documentclass[lettersize,journal]{IEEEtran}
\RequirePackage[table,dvipsnames]{xcolor}
\usepackage[table,dvipsnames]{xcolor}
\usepackage{amsmath,amsfonts}
\usepackage{array}
\usepackage[caption=false,font=normalsize,labelfont=sf,textfont=sf]{subfig}
\usepackage{textcomp}
\usepackage{stfloats}
\usepackage{url}
\usepackage{verbatim}
\usepackage{graphicx}
\usepackage{adjustbox}
\usepackage{arydshln}
\usepackage{cite}
\usepackage{balance}

\usepackage{amsthm}     
\newtheorem{thm}{Theorem}

\usepackage[ruled,vlined,linesnumbered]{algorithm2e} 
\usepackage{amsmath}
\usepackage{enumitem}

\usepackage{colortbl}        
\usepackage{multirow}
\usepackage{pifont}

\usepackage{booktabs}
\usepackage{multirow}

\usepackage{booktabs}
\usepackage{multirow}
\usepackage{makecell}
\usepackage{graphicx}
\usepackage{array}

\usepackage{enumitem}

\usepackage{pifont} 

\makeatletter
\newcommand{\thickhline}{%
    \noalign {\ifnum 0=`}\fi \hrule height 1pt
    \futurelet \reserved@a \@xhline
}
\makeatother

\usepackage{graphicx}
\usepackage{array}
\usepackage{multirow}
\usepackage{amssymb}
\usepackage{booktabs}
\usepackage{caption}

\newtheorem{assumption}[thm]{Assumption}

\definecolor{DarkBlue}{RGB}{64,101,149}

\usepackage{hyperref}

\definecolor{cvprblue}{rgb}{0.0, 0.443, 0.737} 

\hypersetup{
    colorlinks=true,
    citecolor=cvprblue,      
    linkcolor=black,         
    urlcolor=black,           
    filecolor=magenta,       
}

\begin{document}

\title{Toward Polymorphic Backdoor against Semantic Communication via Intensity-Based Poisoning} 




\author{Xiao Yang,
Yuni Lai,
Gaolei Li,
Jun Wu,
Kai Zhou,
Jianhua Li,
and Mingzhe Chen

\thanks{
Xiao Yang, Gaolei Li, Jun Wu, and Jianhua Li are with the School of Computer Science, Shanghai Jiao Tong University, and Shanghai Key Laboratory of Integrated Administration Technologies for Information Security, Shanghai, China (\textit{e-mail}:
\texttt{\{youngshall, gaolei\_li, junwuhn, lijh888\}@sjtu.edu.cn}).

Yuni Lai and Kai Zhou are with the Department of Computing, Hong Kong Polytechnic University, Hung Hom, Kowloon, Hong Kong (\textit{e-mail}:
\texttt{\{yunie.lai@connect.polyu.hk; kaizhou@polyu.edu.hk\}}).

Mingzhe Chen is with the Department of Electrical and Computer Engineering and Frost Institute for Data Science and Computing,
University of Miami, Coral Gables, Florida 33146, USA (\textit{e-mail}: \texttt{mingzhe.chen@miami.edu}).
}
}

\markboth{IEEE Transactions on Information Forensics and Security,~Vol.~**, No.~*, ****~20**}%
{Shell \MakeLowercase{\textit{et al.}}: A Sample Article Using IEEEtran.cls for IEEE Journals}


\maketitle

\begin{abstract}
Semantic Communication (SC) backdoor attacks aim to utilize triggers to manipulate the system into producing predetermined outputs via backdoored shared knowledge.
Current SC backdoors adopt monomorphic paradigms with single attack target, which suffers from limited attack diversity,  efficiency, and flexibility in heterogeneous downstream scenarios.
To overcome the limitations, we propose \emph{SemBugger}, a polymorphic SC backdoor.
By dynamically adjusting the trigger intensity, \emph{SemBugger} finely-grained controls over the SC knowledge to generate diverse malicious results from the system.
Specifically, \emph{SemBugger} is realized through a multi-effect poisoning-training framework.
It introduces graded-intensity triggers to poison training data and optimizes SC systems with hierarchical malicious loss.
The trained system's knowledge dynamically adapts to trigger intensity in inputs to yield target outputs, all while preserving transmission fidelity for benign samples. 
Moreover, to augment SC security, we propose a provable robustness defense that resists \emph{SemBugger}'s homogeneous attacks through a controlled noise mechanism. 
It operates via strategically adding noise in SC inputs, and we formally provide a theoretical lower bound on the defense efficacy.
Experiments across diverse SC models and benchmark datasets indicate that \emph{SemBugger} attains high attack efficacy while maintaining the regular functionality of SC systems. Meanwhile, the designed defense effectively neutralizes \emph{SemBugger} attacks.
\end{abstract}

\begin{IEEEkeywords}
Semantic communication, backdoor attack, polymorphic attack, backdoor defense, data poisoning.
\end{IEEEkeywords}

\section{Introduction}
\IEEEPARstart{D}{iverging} from classical Shannon transmission models, Semantic Communication (SC) establishes a new paradigm, where it processes and transmits semantic units of information rather than conventional raw data bitstreams \cite{9679803, 9955312, 10554663, 10855638, 11395618}.
SC employs shared knowledge or cognitive semantic mappings to selectively extract and convey task-relevant details, and attains remarkable bandwidth reduction.
This knowledge-driven approach supports adaptive compression that dynamically adjusts to both the receiver's knowledge and the specific communication tasks.
Moreover, owing to the bandwidth efficiency and context-aware comprehending capabilities, SC is implemented in several frontier applications, \textit{e.g.}, Extended Reality (XR), Vehicle-to-Everything (V2X), Smart Internet of Things (Smart IoT), and Intelligent City \cite{10198383, 11400632, 10695151, 11027321, 11030612}.

\begin{figure}[t]
    \centering
    \includegraphics[scale=0.52]{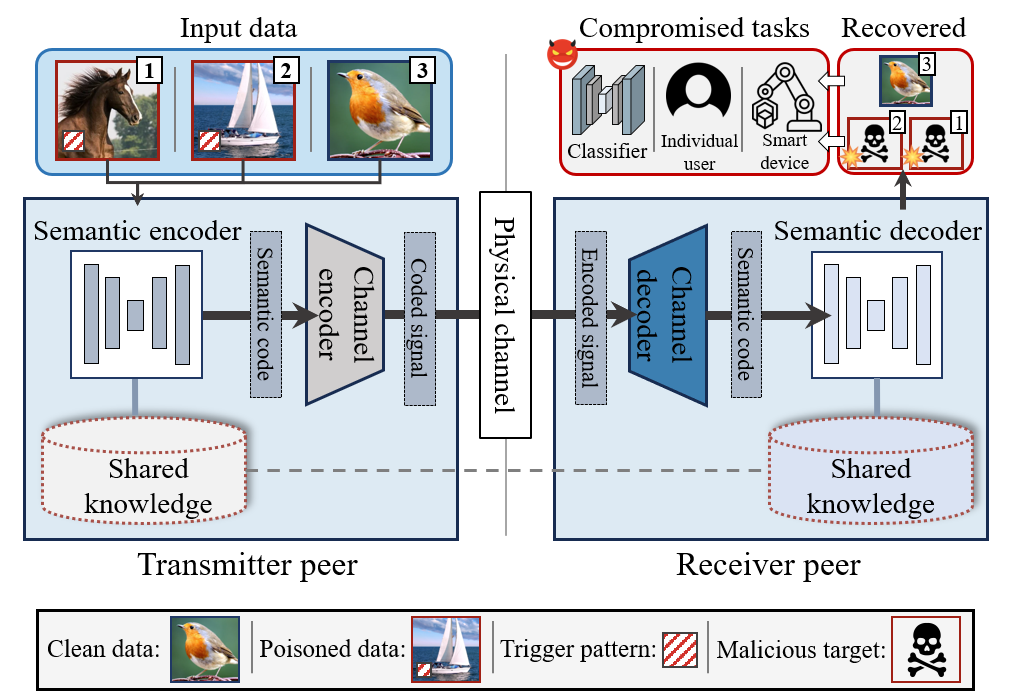}
    \caption{\textbf{Illustration of backdoor attacks against SC systems}. Adversaries embed specific triggers into the input samples at the transmitter side (\textit{i.e.}, data poisoning), inducing the system to deliver poisoned data toward predetermined malicious outputs while retaining normal transmission efficacy for benign samples. These hostile outputs may cause abnormal execution results in downstream tasks and undermine system integrity.}
    \label{intro fig}
\end{figure}

Despite the recent advances in SC, empirical research has uncovered its security vulnerabilities of backdoor threats \cite{10089692, 10622193, 10901411, 10598360, 10744415}.
The illustration of SC backdoors is depicted in Fig.~\ref{intro fig}.
By poisoning the training of SC systems, adversaries can implant malicious functionality in the shared knowledge (\textit{i.e.}, backdoor), which causes the trained system to exhibit predetermined adversarial behaviors whenever it encounters trigger-embedded inputs, while preserving its performance on benign data.
SC backdoors were first investigated in SC downstream tasks, where block-patch triggers were pasted into inputs to induce misclassification \cite{10089692}.
Several works transplanted the attacks into the transmission task, wherein they optimized the trigger design into implicit styles and markedly enhanced backdoor stealthiness \cite{10622193}.
Prior studies have also implanted backdoors into SC-driven autonomous driving systems and manipulated the perception-to-decision transmission pipeline to discard specific visual cues (\textit{i.e.}, triggers), whereby it potentially inducing spurious inferences \cite{10901411}.


Current research can be primarily featured as a monomorphic attack paradigm, wherein the encoding-decoding backdoor mechanism implanted in shared knowledge responds exclusively to a specific trigger or trigger pattern, producing a deterministic, singular malicious output \cite{9802938, 10.1145/3704725, 11040037, 10423783}.
To our knowledge, all existing SC backdoors possess this property.

While prior works have advanced the field, existing backdoor methods adhere to monomorphic paradigms.
This static is constrained to single-objective adversarial manipulations (\textit{i.e.}, transmitting all trigger-embedded data as a fixed malicious target).
However, communication systems inherently operate in “\textit{one-to-many}” or “\textit{many-to-many}” configurations that render adversaries' monolithic attack patterns fundamentally incompatible with the polymorphic requirements of targeted malicious intentionality across diverse users.
Additionally, this paradigm incurs critically deficient flexibility,
and adversaries must implant distinct backdoors tailored to specific malicious goals to fulfill varied intents, which exposes diminished operational efficiency and increases detection risks.






To overcome the above shortcomings, we propose a polymorphic SC backdoor-{\textbf{\emph{SemBugger}}}, \textit{which induces targeted output control via implicit trigger injection into inputs, and achieves differentiable regulation over output results via trigger intensity}.
To implant \emph{SemBugger} backdoor into SC shared knowledge through training for joint source-channel coding, a two-step multi-effect poisoning-training framework is developed. 
It first dynamically synthesizes sample-specific triggers via a neural generator, then poisons the training data by an intensity-stratified injection strategy.
The poisoned data are subsequently applied for SC system training and backdoor implantation, with a hierarchical loss optimizing multi-target attack objectives, regular system functionality, and trigger invisibility simultaneously.
Beyond adversarial investigation, to effectively defend against \emph{SemBugger} attacks, we correspondingly devise a provable robustness defense strategy. With inserting carefully calibrated noise signals into inputs, the proposed strategy neutralizes potential backdoor triggers without compromising systematic functionality for benign data transmission.
Crucially, we formally prove a theoretical lower-bound guarantee on its defensive performance.

Our principal contributions can be summarized below.
\begin{enumerate}[label=$\arabic*$.]
    \item We introduce a polymorphic SC backdoor methodology-\emph{\textbf{\emph{SemBugger}}}. It engenders malicious outputs via shared knowledge from SC systems by implicitly injecting input-specific trigger patterns, where the attack results can be dynamically regulated by trigger intensity.
    Moreover, to implant backdoors into knowledge of victim systems, a multi-effect poisoning-training framework is designed.
    
    \item To mitigate \emph{SemBugger} attacks, a defensive strategy is developed that perturbs inputs with carefully crafted noise. 
    This operation disrupts latent trigger patterns within inputs to prevent backdoor activation and retains minimal functionality impact for benign inputs.

    \item We establish rigorous theoretical guarantees (\textit{i.e.}, certified robustness) for the proposed defense, proving a strict lower bound on its protective efficacy. \textit{To the best of our knowledge, this is the first work that explores certified robustness for SC backdoors}.
    
    \item Experimental evaluations on SC architectures and benchmark datasets manifest that \emph{SemBugger} effectively attains high attack success rates and preserves benign data fidelity. Additionally, the proposed defense strategy efficiently deters potential \emph{SemBugger} threats.
\end{enumerate}

\section{Background and Related Work}
\subsection{Semantic Communication Framework}
The SC system $f:\mathcal{X} \rightarrow \mathcal{X}$ functions by encoding and conveying the semantic information of data.

In this communication architecture, the semantic encoder at the transmitter, denoted as $\mathcal{P}_{\text{enc}}: \mathcal{X} \rightarrow \mathcal{S}$, conducts feature extraction and compression on the source data $x \in \mathcal{X}$ and produces a condensed semantic representation $s = \mathcal{P}_{\text{enc}}(x)$. The Compression Rate $CR$ is defined as the ratio of the original data size to the semantic representation size ($CR = \frac{|x|}{|s|}$).
Next, the system performs channel encoding, which integrates error correction codes and additional redundancy to generate the encoded signal $c = \mathcal{H}_{\text{enc}}(s)$ for transmission:
\begin{align}
\mathbf{X} = \frac{\mathcal{H}_{\mathrm{enc}} \left( \mathcal{P}_{\mathrm{enc}}({x}) \right)}
{\left\| \mathcal{H}_{\mathrm{enc}} \left( \mathcal{P}_{\mathrm{enc}}({x}) \right) \right\|_2},
\quad c' = \mathbf{H} \mathbf{X} + \mathbf{N},
\end{align}
where $\mathbf{H}$ is the channel transfer matrix and $\mathbf{N}$ is the noise vector.
At the receiver side, channel decoding is executed to restore the distorted semantic representation $\hat{s} = \mathcal{H}^{-1}_{\text{enc}}(c')$, where $c'$ represents the received signal affected by channel disturbances like noise and fading.
Finally, the semantic decoder at the receiver, $\mathcal{P}^{-1}_{\text{enc}} : \mathcal{S} \rightarrow \hat{\mathcal{X}}$, processes the recovered semantic data $\hat{s}$ to reconstruct the estimated output $\hat{x} = \mathcal{P}^{-1}_{\text{enc}}(\hat{s})$.
During the training phase, the semantic encoder $\mathcal{P}_{\text{enc}}$ and decoder $\mathcal{P}^{-1}_{\text{enc}}$ are jointly optimized to acquire a shared knowledge base. It has three key aspects: $1)$ extraction of essential semantic features, $2)$ efficient encoding of semantic information, and $3)$ accurate mapping to desired outputs. This guarantees consistent and aligned semantic understanding between communication endpoints \cite{8723589, 9066966, 9998051, 11223146, Zhang_2023_CVPR}.


Besides the above investigations, State-of-the-Art (SOTA) SC research focuses on:
$1)$ Task-Oriented Efficient Coding: Domain-specific semantic encoders are designed to extract mission-critical features (\textit{e.g.}, tumor characteristics in medical diagnostics or navigation-relevant objects in autonomous vehicles) while eliminating non-essential information \cite{9830752}.
$2)$ Joint Semantic-Channel Optimization: This approach synergizes robust semantic symbol construction with adaptive transmission strategies to mitigate time-varying channel distortions \cite{10584091}.
$3)$ Semantic-Centric Resource Management: Network resources are intelligently allocated via semantic importance, which preferentially supports time-critical semantic streams (\textit{e.g.}, emergency notifications) through adaptive bandwidth allocation and edge resource orchestration \cite{9763856}.
$4)$ Interference-Robust Semantic Delivery: 
Reliability under adverse propagation conditions is fortified via salient feature preserving encoding and jointly optimized semantic and channel processing \cite{10460429}.

	
\subsection{Semantic Communication Backdoor}
In SC system \( f \), the adversary $\mathcal{A}$ injects a specific trigger pattern \( \Delta \) into the original input \( x \in \mathcal{X} \) at the transmitter side to construct an adversarial poisoned sample \( x_{\text{adv}} = x \oplus \Delta \), whereby it causes the receiver's decoder \( \mathcal{P}^{-1}_{\text{enc}} \) to produce an $\mathcal{A}$-specified erroneous output \( x_{\text{error}} \neq x \).


\begin{table}[t]
  \centering
  \caption{\textbf{Comparison of SC backdoor} methods in three aspects: $1)$ attack stealthiness, $2)$ generalizability in transmission scenarios, and $3)$ polymorphism for varying targets.}
  \resizebox{\linewidth}{!}{
    \begin{tabular}{c||ccccc}
     \hline\thickhline
     \rowcolor{black!10} \multicolumn{1}{c||}{} & \multicolumn{5}{c}{\textbf{Backdoor Scheme against SC Systems}} \\
     \rowcolor{black!10}
     \multirow{-2}{*}{\textbf{Capacity}} & \textbf{SC Trojan \cite{10089692}} & \textbf{BASS \cite{10622193}} 
                                  & \textbf{IHTG \cite{10901411}} & \textbf{CSBA \cite{10598360}} & \textbf{Ours$^*$} \\
     \hline\hline
     Stealthiness & \ding{55} & \ding{55} & $\checkmark$ & $\checkmark$ & $\checkmark$ \\
     \rowcolor{gray!10} Generality  & \ding{55} & $\checkmark$ & $\checkmark$ & $\checkmark$ & $\checkmark$ \\
     Polymorphism  & \ding{55} & \ding{55} & \ding{55} & \ding{55} & $\checkmark$ \\
     \hline\thickhline
    \end{tabular}
  }
  \label{sc-bkd-compare}
\end{table}

Study \cite{10089692} first explored backdoor attacks in SC systems \( f \). Based on downstream classification, it proposed a basic backdoor method that: $1)$ embeds designed block-patch triggers into selected training samples, and $2)$ reformulates the loss to jointly optimize both the trigger-target association and normal communication performance. 
This dual-optimization preserves regular \( f \) performance while enabling backdoor activation.
Zhou et al. \cite{10622193, 10901411} advanced SC backdoors by extending the applicability from classifications to general transmission scenarios. They
$1)$ enhanced trigger optimization, augmenting the stealthiness of triggering patterns, and
$2)$ redesigned learning loss and kept \( f \) retain normal efficacy and output malicious transmission targets for the poisoned.
Xu et al. \cite{10598360} empirically demonstrated backdoor attacks in autonomous driving SC systems. 
By exploiting specific semantic element as the trigger, their method results in element suppression (\textit{i.e.}, removal) when detected, which realizes covert manipulation via semantic-level backdoor.

Subject to compliance requirements, backdoors can also be leveraged for benign and controllable functionalities \cite{9802938}. $1$) Trigger patterns may be used to implement model watermarking and provenance verification, empowering one to confirm whether a model belongs to a particular vendor and whether it has been stolen or illegally redistributed. $2$) Such mechanisms can further support access control by allowing the model to operate normally only when presented with valid trigger keys, consequently reducing the value of unauthorized deployments. $3$) In security engineering, they may be adapted to honeypot-style practices: embedding known triggers in a controlled environment to systematically evaluate detection capabilities and identify potential attacks.

While preliminary inspections for SC backdoors were conducted, 
\textit{current methodologies remain constrained by a fixed attack target and singular trigger-malicious activation correlation}, with fundamental limitations manifested in operational flexibility. 
Although a few multi-objective studies in Computer Vision (CV) address related problems, the methods designed for classification backdoors (which merely nudge low-dimensional outputs across a decision boundary) are ill-suited to SC, because in this setting, a backdoor must concurrently $1)$ manipulate high-dimensional continuous reconstructions, $2)$ remain robust to channel noise and codec compression, and $3)$ preserve normal semantic fidelity under the tightly coupled parameters of transmitter and receiver. These requirements render current methods ineffective for SC \cite{9211729, WANG2024110449, 9551983, 10901170}.

To better summarize related research, Tab. \ref{sc-bkd-compare} compares SC backdoors in three aspects: $1)$ backdoor stealthiness, $2)$ generalizability for transmission tasks, and $3)$ polymorphism to varying malicious targets.

\subsection{Backdoor Threat Defense}
Current neural backdoor defenses can primarily be categorized via four aspects: \emph{data}, \emph{training}, \emph{model}, and \emph{deployment} \cite{9802938, 9772337}.
$1$) Regarding data side, suspicious poisoning can be identified via anomalous-sample consistency checks, together with establishing a traceable data supply chain. $2$) Concerning training side, robust training and regularization, noise injection, and trigger-suspicion–driven training constraints can be adopted to reduce the model overfitting to rare trigger patterns. 
$3$) For model side, backdoor representations can be weakened or removed via neuron pruning and channel sparsification, reverse-engineering–based trigger search and sanitization, and fine-tuning with clean data. 
$4$) From deployment side, combining input anomaly detection and output consistency monitoring mechanisms powers degradation or blocking when suspected trigger behaviors are detected.

Although backdoor defenses for classification tasks have been extensively studied, defensive strategies for transmission tasks (\textit{i.e.}, SC) remain underexplored. To the best of our knowledge, there are currently no defense methods specifically designed for backdoors in SC systems.

\section{Proposed Methodology}
We first present the design requirements and then formalize the details of our proposed \emph{SemBugger} framework.

\begin{figure*}[t] 
    \centering
    \includegraphics[scale=0.7]{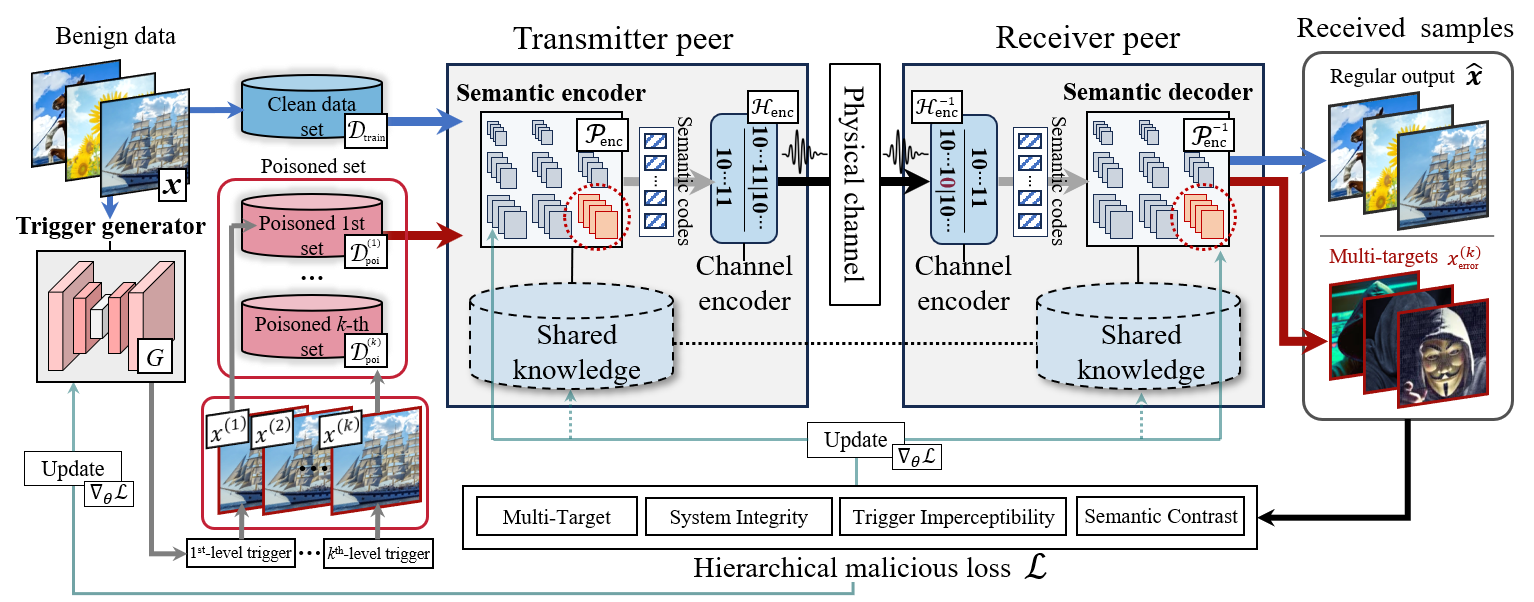} 
    \caption[Concise figure description]{
        \textbf{Illustration of the proposed \emph{SemBugger}}. $1)$ Selected victim samples are injected with multi-intensity triggers crafted by the trigger generator to construct a multi-dimensional poisoned dataset. $2)$ By training with hierarchical loss, the system learns to transform inputs with varying-intensity triggers into differentiated malicious targets, while preserving functionality on benign data. $3)$ Adversaries achieve differential control over hostile outputs by embedding multi-intensity triggers in inputs.
    }
    \label{framework}
\end{figure*}

\subsection{Design Requirements}
\label{design_require}
The present research seeks to develop a backdoor approach characterized by three foremost features:

\vspace{0.8ex}
\noindent \textbf{Flexibility}. 
The proposed attack framework should require robust polymorphic target compatibility while constructing a tailored control scheme for heterogeneous attack targets.
Consequently, it will increase attack versatility.

\vspace{0.8ex}
\noindent \textbf{Stealthiness}. 
The backdoored SC system should sustain uncompromised robust functionality across multiple transmission environments.
This preserves the integrity of original systems, retaining uninterrupted service availability and indirectly raising the backdoor's stealthiness.

\vspace{0.8ex}
\noindent \textbf{Imperceptibility}. 
The attack data must remain visually and statistically indistinguishable from legitimate data to circumvent human analyst inspections. This covertness directly contributes to adversarial efficacy by reducing anomaly alert during manual triage, and increasing the mean time to detection.

\subsection{Methodology Insights}
\label{insight}
To fulfill the method criteria in Sec. \ref{design_require}, we embed imperceptible trigger noise within input samples to enable precise modulation of backdoor activation dynamics.
$1)$ Through precise modulation of noise ratio, we attain multi-target control over backdoor activation outputs and eliminate external control variables.
Concurrently, throughout the optimization procedure, $2)$ we preserve the original system's training performance metrics to secure unimpaired transmission proficiency.
$3)$ The amplitude of the noise is optimized to persist that the trigger-embedded samples display no perceptible anomalies. 
Also, $4)$ contrary to conventional spatially localized trigger patterns, the noise is globally distributed, whereby it demonstrates better spatial uniformity and robust partial-trigger activation capabilities.


\subsection{{\textbf{SemBugger}} Framework}
\label{method framework}
With the insights from Sec. \ref{insight}, the implementation details of our proposed \emph{SemBugger} are given.
We adopt a multi-effect poisoning-training framework to implant backdoors in SC system $f$'s shared knowledge, whereby it enables controlled generation of multiple malicious transmission results during deployment by precise regulation of trigger noise ratio.

The poisoning-training framework consists of two steps: $1)$ Multi-Dimensional Data Poisoning and $2)$ Hierarchical System Backdoor Training.
In the $1$st step, a trigger generator is deployed to infect the training dataset $\mathcal{D}_\text{train}$ with precisely calibrated noise ratios.
Subsequently, $f$ performs system training using a hybrid dataset comprising both poisoned set $\mathcal{D}_\text{poi}$ and clean data $\mathcal{D}_\text{train}$ to embed our desired backdoor functionality.
The training procedure is schematically illustrated in Fig. \ref{framework}.

\subsubsection{Multi-Dimensional Data Poisoning} 
The available clean training dataset $\mathcal{D}_\text{train}$ is applied to synthesize a poisoned dataset $\mathcal{D}_\text{poi}$ through a neural trigger generator $\mathcal{G}$. We suppose $T$ different adversarial targets will be configured in attacks.

Specifically, based on the poisoning rate parameter $\gamma \in (0,1)$, we first construct the victim data $\mathcal{D}_\text{victim}$ by randomly sampling from the original training set $\mathcal{D}_\text{train}$ via sampling function $\pi_\gamma: \mathcal{D}_\text{train} \to \mathcal{D}_\text{victim}$, such that $|\mathcal{D}_\text{victim}| = \gamma|\mathcal{D}_\text{train}|$:
\begin{equation}
\mathcal{D}_\text{victim} = \left\{ x_i \mid x_i \sim \pi_\gamma(\mathcal{D}_\text{train}),\ i=1,\dots,\lfloor \gamma N \rfloor \right\},
\end{equation}
where $N = |\mathcal{D}_\text{train}|$ denotes the cardinality of the training set.

With $\mathcal{D}_\text{victim}$, for each sample $x_i$ in it, a sample-specific trigger $\Delta_i$ is synthesized via a neural generator $\mathcal{G}$ (we adopt an Attention U-Net encoder architecture \cite{oktay2018attentionunetlearninglook} for $\mathcal{G}$'s setup):
\begin{equation}
    \Delta_i = \mathcal{G}(x_i), \quad \text{where} \quad x_i \in \mathcal{D}_{\text{victim}}.
\end{equation}
For $T$ different attack targets, we construct $T$ poisoned subsets, where the $k$-th subset $\mathcal{D}^{(k)}_{\text{poi}}$ contains all poisonous samples with trigger noise applied at a ratio of $k/T$ (we define it as the $k$-th level trigger noise):  
\begin{equation}
    \mathcal{D}^{(k)}_{\text{poi}} = \left\{ x_i + \frac{k}{T} \Delta_i \;\middle|\; x_i \in \mathcal{D}_{\text{victim}} \right\}, \quad \forall k \in \{1, 2, \ \cdots, T\}.
\end{equation}
Finally, the complete poisoned dataset $\mathcal{D}_{\text{poi}}$ is the union of these subsets $\mathcal{D}^{(k)}_{\text{poi}}$, which is expressed as  
\begin{equation}
\mathcal{D}_{\text{poi}} = \bigcup_{k=1}^{T} \mathcal{D}^{(k)}_{\text{poi}}.
\end{equation}

The poisoned data $\mathcal{D}_{\text{poi}}$ and trigger generator $\mathcal{G}$ will be exploited in system $f$ training for backdoor implantation.

\subsubsection{Hierarchical System Backdoor Training} 
By training, we embed backdoor behaviors into $f$, and force it to generate specified malicious transmission outputs according to varying trigger noise ratios present in inputs.

Specifically, we take a hierarchical malicious loss-based training framework for system $f$, which jointly optimizes $f$ via the clean training set $\mathcal{D}_{\text{train}}$ and poisoned dataset $\mathcal{D}_{\text{poi}}$ to derive controlled backdoor implantation. 
First, by minimizing the feature distance between poisoned samples $\mathcal{D}_{\text{poi}}$ and target malicious outputs, the trigger $\Delta$ is rigorously constrained to divert input samples toward predetermined mistransmission of $f$.
Second, $f$'s regular performance on benign data is maintained by optimizing the standard transmission loss function over clean set $\mathcal{D}_{\text{train}}$.
Moreover, the imperceptibility of $\Delta$ is augmented by restricting the perceptual similarity between poisoned and original source samples in the visual domain.
Gaining from the above analysis, this multi-tiered optimization architecture is expressed through training SC system with the following loss functions.

\vspace{0.8ex}
\noindent \textbf{Multi-Target Loss} \textbf{$\mathcal{L}_{a}$}: We promote polymorphic backdoor implantation for regulable multi-target attacks in SC systems:
\begin{equation}
    \mathcal{L}_a = \sum_{k=1}^T \mathbb{E}_{x^{(k)}_{i} \in \mathcal{D}^{(k)}_{\text{poi}}} \left[ 
        \underbrace{\| x_{\text{error}}^{(k)} - f(x^{(k)}_{i}, \theta) \|_2^2}_{{\scriptsize \textit{Attack} \ \textit{Optimization}}} 
    \right],
\end{equation}
where $x_{\text{error}}^{(k)}$ denotes the $k$-th malicious transmission target specified by adversaries and $\theta$ implies the semantic encoder-decoder parameters.
Through optimizing the loss between $f(x^{(k)}_{i}, \theta)$ and $k$-th manipulated target $x_{\text{error}}^{(k)}$, the trained $f$ system will transit data containing $k$-th level triggers as $x_{\text{error}}^{(k)}$ upon deployment, since $f$ learns shared knowledge of mapping $k$-th level triggers to $x_{\text{error}}^{(k)}$.

\vspace{0.8ex}
\noindent \textbf{System Integrity Loss} \textbf{$\mathcal{L}_{b}$}: It upholds the functional integrity of benign sample transmission in the SC system:
\begin{equation}
    \mathcal{L}_b = \mathbb{E}_{x_{i} \in \mathcal{D}_{\text{train}}} \left[ \| x_{i} - f(x_{i}, \theta) \|_2^2 \right].
\end{equation}
By minimizing the loss between $f(x_{i}, \theta)$ and benign samples $x_{i}$, the trained $f$ system preserves its data transmission capability for normal (\textit{i.e.}, unperturbed) data.



\vspace{0.8ex}
\noindent \textbf{Trigger Imperceptibility Loss} \textbf{$\mathcal{L}_{p}$}: Trigger perceptibility is minimized through optimization, leading to augmented attack covertness for trigger embedding:
\begin{equation}
    \begin{split}
        \mathcal{L}_{p} &= \sum_{k=1}^T \mathbb{E}_{x^{(k)}_{i} \in \mathcal{D}^{(k)}_{\text{poi}}, x_{i} \sim \mathcal{D}_{\text{train}}} \Big[ w_1 \left(1  - \text{SSIM}(x_{i}, x^{(k)}_{i}) \right) \\
        &\quad + w_2 \left(1 - \text{TC}(x_{i}, x^{(k)}_{i}) \right) + w_3 \left(1 - \text{CSIM}(x_{i}, x^{(k)}_{i}) \right) \Big],  
    \end{split}
\end{equation}
where $\text{SSIM}(\cdot,\cdot)$ symbolizes Structural Similarity Index Measure; $\text{TC}(\cdot,\cdot)$ represents Tanimoto Coefficient; and $\text{CSIM}(\cdot,\cdot)$ stands for Cosine Similarity.
With the similarity (\textit{i.e.}, $\text{SSIM}$, $\text{TC}$, and $\text{CSIM}$) between $x^{(k)}_{i}$ and $x_{i}$ refined, the trigger generator $\mathcal{G}$ learns to craft imperceptible implicit triggers $\Delta$.

\vspace{0.8ex}
\noindent \textbf{Semantic Contrastive Loss} \textbf{$\mathcal{L}_{c}$}: We separate the semantic representations of poisoned samples and benign samples to accelerate SC system training convergence:
\begin{equation}
    \begin{split}
        \mathcal{L}_{c} &= \sum_{k=1}^T \mathbb{E}_{x^{(k)}_{i} \in \mathcal{D}^{(k)}_{\text{poi}}, x_{i} \sim \mathcal{D}_{\text{train}}} \\
        &\quad \Big[ \max \left(0,  m - \text{Dis}\left( \mathcal{P}_{\text{enc}}(x_{i}),\ \mathcal{P}_{\text{enc}}(x^{(k)}_{i}) \right)  \right) \Big].
    \end{split}
\end{equation}
The optimization of distance between $\mathcal{P}_{\text{enc}}(x_{i})$ and $\mathcal{P}_{\text{enc}}(x^{(k)}_{i})$ induces discriminative latent space separation for infected and normal samples after semantic encoding, which serves as accelerating convergence in training.

\vspace{0.8ex}
\noindent \textbf{Overall Training Loss}.  
Combining the above analytical results, we formulate the final loss optimization:
\begin{equation}
\label{total loss}
\mathcal{L} =  \lambda_a\cdot\mathcal{L}_a + \lambda_b\cdot\mathcal{L}_b + 
\lambda_p\cdot\mathcal{L}_p +
\lambda_c\cdot\mathcal{L}_c, 
\end{equation}
\begin{equation}
\left\{
\begin{aligned}
\theta^{(t+1)} &\leftarrow \theta^{(t)} - \eta_\theta \cdot \nabla_\theta {\mathcal{L}}_\theta, \\
\theta^{(t+1)}_\mathcal{G} &\leftarrow \theta^{(t)}_\mathcal{G} - \eta_{\mathcal{G}} \cdot \nabla_{\theta_\mathcal{G}} {\mathcal{L}}_{\theta_\mathcal{G}},
\end{aligned}
\right.
\end{equation}
where $\theta_\mathcal{G}$ symbolizes parameters of the trigger generator $\mathcal{G}$.
Through the training, we complete backdoor implantation within the SC system. The algorithmic framework of \emph{SemBugger} is formally presented in Alg. \ref{alg-atk}.

\begin{algorithm}[t]
\caption{\emph{SemBugger}}
\label{alg-atk}
\KwIn{Initial SC system $f(\cdot,\cdot)$ (w/ para. $\theta$), Trigger generator $\mathcal{G}(\cdot)$ (w/ para. $\theta_\mathcal{G}$), Poisoning rate $\gamma$, Malicious multi-targets $\big\{ x_{\text{error}}^{(k)} \big\}_{k=1}^T$, Training dataset $\mathcal{D}_\text{train}$, and Training epoch amount \textit{M}}
\KwOut{Backdoored SC system $f(\cdot)$ and Trained trigger generator $\mathcal{G}(\cdot)$}
\vspace{0.07cm}
{\footnotesize{\color{DarkBlue}{\tcc{Victim Data Selection}}}}

$\mathcal{D}_{\text{victim}} \sim_{\gamma} \mathcal{D}_\text{train}$ \\
\ForEach{epoch $\leftarrow$ 1 \ to \ M}{
    $\left\{ \mathcal{D}^{(k)}_{\text{poi}} \right\}_{k=1}^T \gets \emptyset$ \\
    {\footnotesize{\color{DarkBlue}{\tcc{Multi-Dimensional Data Poisoning}}}}
    \ForEach{sample $x_i \in \mathcal{D}_{\text{victim}}$}{
        $\Delta_i = \mathcal{G} \left( x_i \right)$ \\
        \ForEach{k $\leftarrow$ 1 \ to \ T}{
            $\mathcal{D}^{(k)}_{\text{poi}} \leftarrow \mathcal{D}^{(k)}_{\text{poi}} \cup \left\{ x_i + \frac{k}{T} \Delta_i \right\}$
        }
    }
    $\mathcal{D}_{\text{poi}} = \bigcup_{k=1}^{T} \mathcal{D}^{(k)}_{\text{poi}}$
    
    {\footnotesize{\color{DarkBlue}{\tcc{Hierarchical System Backdoor Training}}}}

    \ForEach{sample $x_i \in \mathcal{D}_{\text{train}}$}{
        Compute $f(x_{i}, \theta)$
    }

    \ForEach{$\mathcal{D}^{(k)}_{\text{poi}} \in \mathcal{D}_{\text{poi}}$}{
    \ForEach{sample $x^{(k)}_{i} \in \mathcal{D}^{(k)}_{\text{poi}}$}{
        Compute $f(x^{(k)}_{i}, \theta)$
    }}

        Compute loss $\mathcal{L}_{a}$, $\mathcal{L}_{b}$, $\mathcal{L}_{p}$, and $\mathcal{L}_{c}$ in Eq. \eqref{total loss}\\
        $\theta = \theta  - \eta \nabla_{\theta} \mathcal{L}$;
        $\theta_\mathcal{G} = \theta_\mathcal{G}  - \eta \nabla_{\theta_\mathcal{G}} \mathcal{L}$
}
{\footnotesize{\color{DarkBlue}{\tcc{SC system gets backdoored after training}}}}
\Return $f_(\cdot,\cdot)$ and $\mathcal{G}(\cdot)$\
\end{algorithm}

\subsection{Attack Conducting}
With the completion of SC system $f$ training via the framework in Sec. \ref{method framework}, $f$ (or its shared knowledge) gets backdoored. 
More precisely, after $f$ deployment, adversaries intending to manipulate $f$'s outputs to produce the $k$-th order malicious target $x_{\text{error}}^{(k)}$ (\textit{i.e.}, backdoor activation) can simply inject the corresponding $k$-th level trigger noise $\Delta$ into the input data stream to operationalize diversely regulable SC system reconstructed result governance.

\section{Certified Defense}
Categorically, there are two aspects of defenses to build robust learning systems: \textit{empirical approaches} and \textit{certified methods}. The former class typically targets mitigation of known attack variants but remains vulnerable to sophisticated adaptive adversaries, and it brings about a cat-and-mouse game between adversaries and defenders.
For example, in image classification backdoors, \cite{9797338} introduced a dynamic attack scheme that can bypass SOTA empirical protections \cite{10.1145/3359789.3359790, YangLZLLZ25, 10.1145/3579856.3582822, 8835365}.
Consequently, our work is conducted around certified defense.
In machine learning classification tasks, a certified defense guarantees consistent label prediction for all data points within a specified region around an input. 
With this foundation, we formally define certified backdoor defense in SC systems as maintaining identical transmission outputs for all data instances within a determined input region.
In this section, the threat model is given first, then the defense method is detailed. Finally, we present rigorous theoretical robustness proofs of our defense method.

\subsection{Threat Model}
We take dual-perspective analysis from both the adversarial and defensive considering.

\vspace{0.8ex}
\noindent \textbf{Adversary Assumptions}. 
Our threat model aligns with previous backdoor paradigms (\textit{e.g.}, SC Trojan, BASS, and CSBA) in SC. A compromised SC system, acquired by users through third-party training services or post-training modifications, holds dual behavior: $1)$ maintaining nominal performance on legitimate system inputs, while $2)$ executing adversarial misinterpretations and outputs when encountering trigger-embedded data.
We adopt the strongest \textit{white-box} attack assumption, wherein adversaries possess full knowledge of system parameters, architecture, loss functions, training data, and can leverage auxiliary datasets for attacks.
This setting will stringently evaluate the efficacy of the defense mechanism.

\vspace{0.8ex}
\noindent \textbf{Defender Assumptions}. 
The key capability resides in input data controllability, whereby we build adversarial robustness through preprocessing like noise addition and feature transformation and maintain \textit{black-box} compatibility by avoiding system architecture modifications.
Crucially, the defense provides formally verifiable security guarantees that assure output reliability within specified perturbation bounds. It can fundamentally preserve transmission consistency across both clean and adversarially modified inputs, and optimally balance robustness with system utility to retain regular performance. 

\subsection{Defense Framework}
\label{defense-framework}
\begin{figure}[t]
    \centering
    \includegraphics[scale=0.81]{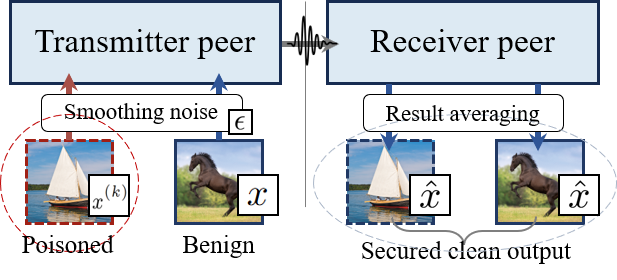}
    \caption{\textbf{Illustration of defense strategy against SemBugger}. It is deployed during the operational phase of the SC system. Before data is input into the system, smoothed noise is added to invalidate potential triggers, which guarantees normal output at the receiver end without affecting regular data transmission.}
    \label{defense_fmwk}
\end{figure}

To mitigate backdoor impacts, we propose a general backdoor defense strategy named \textit{semantic smoothing} that is model-agnostic and training-free.  
Next, we delineate the following aspects: $1)$ the data transmission methodology for testing inputs utilizing a smoothed SC system, $2)$ the theoretical robustness guarantees provided by the smoothed SC transmission systems. The architecture of the proposed defense framework is depicted in Fig. \ref{defense_fmwk}.

\vspace{0.8ex}
\noindent \textbf{Semantic Smoothing}. 
Consider a transmission problem from the input space $\mathbb{R}^d$ to the output space $\mathbb{R}^d$. Semantic Smoothing constructs a smoothed transmitter $S$ from a base telecommute transmitter $f$, where the output of $S(x)$ is determined by averaging the transmission result of $f$ under noise perturbation. We define the smoothed transmitter as follows:

\begin{equation}\label{eq:smoothed_transmitter}
S(x) = \mathbb{E}_{\epsilon \sim \mathcal{N}(0, \sigma^2 I)} \big[f(x + \epsilon)\big],
\end{equation}
where the noise $\epsilon$ follows an isotropic Gaussian distribution $\mathcal{N}(0, \sigma^2 I)$.
The hyperparameter \( \sigma \) controls the noise intensity ($0.25$ for experiments), balancing the {\textit{protection}} and {\textit{efficacy}} of the transmission process. Considering all the image pixels are in $[0,1]$, we have the theorem that the backdoored transmitted image is bounded:
\begin{thm}
\label{thm1}
    (Semantic robustness guarantee) Given the transmission function $f:\mathbb{R}^d\rightarrow\mathbb{R}^d$, random noise $\epsilon \sim \mathcal{N}(0, \sigma^2 I)$, the smoothed transmission function $S:\mathbb{R}^d\rightarrow\mathbb{R}^d$ defined in \eqref{eq:smoothed_transmitter}, and $\forall i=1,2,\cdots, d, f(x)_i\leq 1$, then the smoothed transmitter is $L$-Lipschitz:
    \begin{equation}
        \forall x, x'\in \mathbb{R}^d, ||S(x)-S(x')||_{\infty}\leq L||x-x'||_2,
    \end{equation}
    where the $L=\sqrt{\frac{2}{\pi \sigma^2}}$.  
\end{thm}
The proof for Theorem \ref{thm1} is in the \href{https://github.com/youngshallyx/Toward-Polymorphic-Backdoor-against-Semantic-Communication-via-Intensity-Based-Poisoning-Supp-/blob/main/IEEE%20TIFS-Supplementary%20Materials.pdf}{\emph{Supplemental Material}}. This result guarantees that the transmitted result under attacks is close to the one with clean input, providing the robustness of the smoothed transmit function.

\noindent \textbf{Successful Defense}. For backdoored input $x':=x\oplus\Delta$, and $T$ backdoor targets $err:=\{x_{error}^{(k)}\}_{k=1}^{T}$, we guarantee that the transmit results are not close to any of the targets. To describe it, we define a defense successful judgment function $j(\cdot|err):\mathbb{R}^d\rightarrow\{0,1\}$, where $0$ denotes successful defense (correct transmission) and $1$ denotes successful attack (error transmission). Specifically, for each attack input containing triggers, we measure the SSIM between the reconstruction result (\textit{i.e.}, SC output) and the predefined malicious target: $\text{SSIM}(f(x'),x_{error}^{(k)})$. A valid attack is registered when the computed SSIM value surpasses the threshold $\tau$, where the $\tau$ is determined via statistical analysis of the SSIM values collected under normal (attack-free) input $\text{SSIM}(f(x),x_{error}^{(k)})$. Specifically, we first compute the mean $\mu$ and standard deviation $\sigma$ of $\text{SSIM}(f(x),x_{error}^{(k)})$ values from $1000$ benign communication samples. $\tau$ is then established as:
$\tau = \mu - 3\sigma$. Then, $j(f(x')|err)=1$ if $max \text{ SSIM}(f(x'),x_{error}^{(k)})\geq\tau$, otherwise $j(f(x')|err)=0$. 

We define the smoothed defense successful judgment function as follows:
\begin{equation}\label{eq:smoothed_judge}
    J(x) = \arg\max_{y\in\{0,1\}} \mathbb{P}\big[j(f(x + \epsilon)|err) = y \big],
    \end{equation}
where $\epsilon \sim \mathcal{N}(0, \sigma^2 I)$. We note that when we obtain the smoothed transmission results $S(x)$ defined in \eqref{eq:smoothed_transmitter}, we can obtain $J(x)$ at the same time with negligible computation workload. 
\begin{assumption}
\label{assumption1}
We assume that, given a clean input $x\notin err$, it cannot be backdoored successfully because there is no trigger, then we have $J(x)=0$. 
\end{assumption}

\begin{thm}
\label{thm2}
    (Successful defense guarantee) Given an input $x$, and the smoothed judgment function $J(\cdot)$ defined in \eqref{eq:smoothed_judge}. For attack input $x':=x\oplus\Delta$, we guarantee that $J(x')=0$ (successful defense) if $||x'-x||_2\leq R$, where $R=\sigma(\Phi^{-1}(\underline{p_0}))$, $\Phi^{-1}$ is the inverse of the standard Gaussian cumulative distribution function, and $\underline{p_0}$ is the lower bound of the probability $\mathbb{P}\big[j(f(x + \epsilon)|err) = 0 \big]$. 
\end{thm}
The proof for Theorem \ref{thm2} is in the \href{https://github.com/youngshallyx/Toward-Polymorphic-Backdoor-against-Semantic-Communication-via-Intensity-Based-Poisoning-Supp-/blob/main/IEEE%20TIFS-Supplementary%20Materials.pdf}{\emph{Supplemental Material}}. The results provide the guarantee that we can obtain a successful defense as long as the backdoor images are close to the clean image. 
The algorithm of our proposed certified defense framework is illustrated in Alg. \ref{alg_semantic_smoothing}.

\begin{algorithm}[t]
\caption{\emph{Semantic Smoothing Defense Framework}}
\label{alg_semantic_smoothing}
\KwIn{
Base (possibly backdoored) SC transmission system $f(\cdot;\theta):\mathbb{R}^d\!\rightarrow\!\mathbb{R}^d$,
test input set $\mathcal{D}_{\text{test}}$ (or a single input $x$)
}
\KwOut{
Defended transmission outputs $\{\hat{S}(x)\}_{x\in\mathcal{D}_{\text{test}}}$
}

{\footnotesize{\color{DarkBlue}{\tcc{Transmission Phase Deployment}}}}

\ForEach{$x \in \mathcal{D}_{\text{test}}$}{
    $y_{\text{sum}} \gets \mathbf{0}$

    {\footnotesize{\color{DarkBlue}{\tcc{Transmitter Smoothing}}}}
    \For{$j \leftarrow 1$ \KwTo $N$}{
        Sample $\epsilon_j \sim \mathcal{N}(0,\sigma^2 I)$\\
        $\tilde{x}_j \gets \Pi_{[0,1]}(x + \epsilon_j)$\\
        {\footnotesize{\color{DarkBlue}{\tcc{Peer End Receiving}}}}
        $y_{\text{sum}} \gets y_{\text{sum}} + f(\tilde{x}_j;\theta)$
    }
    {\footnotesize{\color{DarkBlue}{\tcc{Result Constructing}}}}
    $\hat{S}(x) \gets \frac{1}{N}y_{\text{sum}}$ \\
}
\Return Smoothed transmission results $\{\hat{S}(x)\}_{x\in\mathcal{D}_{\text{test}}}$
\end{algorithm}

\vspace{0.8ex}
\noindent \textbf{Practical Implementation}. 
Since our defense relies solely on injecting semantic-smoothing noise into the input samples, it requires no model retraining and introduces no additional modules such as detectors, filters, or auxiliary networks. Therefore, it can be deployed in a plug-and-play manner during the runtime stage. Concretely, $1$) in online transmission, the transmitter applies a smoothed perturbation to each incoming input sample (regardless of whether it is benign or potentially trigger-bearing) according to a predefined noise, and then feeds the perturbed sample into the semantic communication system for transmission. $2$) The receiver subsequently reconstructs the transmission result to obtain a smoothed output according to Eq. \eqref{eq:smoothed_transmitter}. By performing randomized smoothing in the input semantic space, this procedure suppresses backdoor activation while preserving the transmission efficacy on normal data.

\section{Experiment}
We validate the efficacy of the proposed \emph{SemBugger} attack through benchmarking against SOTA baselines. To examine influences of individual conditions, parameter-wise ablation analyses are also conducted.
Additionally, the robustness of our defensive mechanism is empirically verified against the developed \emph{SemBugger} methodology.
We first present the experimental setting and subsequently proceed to the test results.

\subsection{Experimental Settings}

\subsubsection{Victim SC Systems}
$5$ SOTA SC systems were adopted for experimental assessment.
$1)$ JSCC (Joint Source-Channel Coding) utilizes deep convolutional neural networks with residual connections to jointly optimize source compression and channel coding in an end-to-end learnable framework, which removes the demands for separate modular designs \cite{8723589};
$2)$ JSCC-f (Deep Joint Source-Channel Coding with Feedback) introduces a novel two-way feedback mechanism comprising channel state information reporting and acknowledgment signals, whereby it enables real-time adaptive modulation of the encoder's latent representations under time-varying fading channels \cite{9066966};
$3)$ JSCC-q (Constellation Constrained Deep Joint Source-Channel Coding) implements a trainable vector quantization module with learnable codebook embeddings that discretizes the continuous latent space into finite clusters while maintaining gradient flow through straight-through estimator backpropagation \cite{9998051};
$4)$ SCAN (Semantic Channel-Adaptive Networking) employs a content-aware gating mechanism that dynamically adjusts wavelet-based compression ratios by jointly analyzing semantic saliency maps and instantaneous channel capacity metrics to realize optimal rate-distortion tradeoffs \cite{10510413};
$5)$ SemCC (Semantic Contrastive Coding) reformulates channel-induced impairments as differentiable noise layers within a Siamese network, where contrastive loss minimization aligns noisy and clean samples in a shared embedding space to enhance semantic invariance \cite{10530261}.

\subsubsection{Testing Datasets}
This study evaluated method performance across $4$ standard CV benchmarks:
$1)$ MNIST: A collection of $70,000$ handwritten digit images ($0-9$) with $28\times28$ grayscale pixels, featuring perfectly balanced classes ($6,000$ training and $1,000$ test samples per digit) and $11.18\%$ average sparsity \cite{6296535}.
$2)$ F-MNIST: A clothing alternative maintaining MNIST's format ($70$\,k $28\times28$ grayscale images) but containing $10$ fashion item categories (\textit{e.g.}, shirts and sneakers), which indicates greater visual complexity \cite{xiao2017fashionmnistnovelimagedataset}.
$3)$ CIFAR-10: A set of $60,000$ tiny $32\times32$ color images across $10$ object classes (animals, vehicles etc.) with measured $15.7\%$ label noise \cite{8745428}.   
$4)$ ImageNet: The large-scale visual database with $1.28$ million high-resolution natural images (average $469\times387$ pixels) spanning $1,000$ everyday object categories in realistic long-tailed distribution (most frequent class: $3,170$ samples; rarest: $468$) \cite{5206848}.
Our tests harnessed a class-balanced subset containing $5$ randomly sampled categories, with a total of $40,000$ training images and $10,000$ test images ($8,000$ training and $2,000$ testing instances per category).

\subsubsection{Evaluation Metrics}
The method performance was assessed along $2$ primary aspects: efficacy and stealthiness.

\vspace{0.8ex}
\noindent \textbf{Attack Efficacy}. 
We employed the Attack Success Rate (ASR) to evaluate attack capability. Specifically, since our attack method manipulates SC system into producing malicious outputs by injecting triggers into inputs, ASR is calculated as:

\begin{equation}
\text{ASR} = \frac{\text{\# \textit{Successful Attack Trials}}}{\text{\# \textit{Total Number of Attacks}}},
\label{eq-asr}
\end{equation}
where the \textit{Successful Attack Trials} refers to the count of instances where the SC system produces malicious outputs under the influence of the trigger, while the \textit{Total Number of Attacks} represents all attempted attack samples. 
Unless otherwise specified, ASR refers to the average attack success rate across all multi-target attacks.
Specifically, \textit{we utilized $\text{ASR}_i$ to denote ASR for the \textit{i}-th malicious target result (\textit{i.e.}, $x_{\text{error}}^{(k)}$)}. We define the criteria for a successful attack with the attack successful judgment function $j(x|err)$ described in Sec.~\ref{defense-framework}. 
This metric directly reflects the attack effectiveness of the method, and a higher ASR indicates stronger control over victim SC systems. 

\vspace{0.8ex}
\noindent \textbf{Attack Stealthiness}. 
To strengthen the attack stealthiness, it is imperative that the backdoored system preserves the regular functionality with minimal deviation.
Namely, backdoored and benign systems should maintain highly consistent transmission efficiency when processing benign data.
Therefore, the Peak Signal-to-Noise Ratio (PSNR) difference was used to evaluate the backdoor stealthiness. 
We take PSNR$^\circ$ to denote the benign data PSNR of the unattacked SC system, and PSNR* to represent that of the backdoored system.
For our objective, the efficiency gap ($\Delta$PSNR) between PSNR$^\circ$ and PSNR*:
\begin{equation}
    \Delta \text{PSNR} = \text{PSNR}^\circ - \text{PSNR}^*
    \label{eq:delta_psnr},
\end{equation}
ought to be kept as low as possible.

PSNR measures the distortion between a processed signal and the original. 
It is derived from the Mean Squared Error (MSE) and expressed in decibels (dB), and higher values indicate less distortion.
The PSNR formula is:
\begin{equation}
    \text{PSNR} = 10 \cdot \log_{10} \left( \frac{\text{MAX}_I^2}{\text{MSE}} \right),
\end{equation}
where: $\text{MAX}_I$ is the maximum possible pixel value (\textit{e.g.}, $255$ for $8$-bit images) and $\text{MSE}$ (Mean Squared Error) is given by
\begin{equation}
\text{MSE} = \frac{1}{mn} \sum_{i=0}^{m-1} \sum_{j=0}^{n-1} \left[ I(i,j) - K(i,j) \right]^2,
\end{equation}
where $I$ and $K$ represent the original and distorted images, while $m$ and $n$ denote the image resolution.

\subsubsection{Comparison Baselines}
We levered $3$ SOTA SC backdoor methodologies as benchmarks: 
$1)$ SC Trojan, the pioneering backdoor attack framework originally designed for downstream classification tasks, which we adapt to transmission scenarios through loss function modification \cite{10089692}; 
$2)$ BASS, the first dedicated backdoor attack method specifically engineered for SC transmission tasks, capable of precise targeted output induction via patch trigger \cite{10622193}; and 
$3)$ IHTG, an adjusted variant of BASS that considerably improves attack stealthiness through optimized trigger pattern design, with the triggers made substantially more imperceptible \cite{10901411}.

\subsection{Comparison Study}
\label{comparison result}
We evaluated the attack performance of \emph{SemBugger} on $4$ distinct datasets by monitoring both ASR and the $\Delta$PSNR for benign samples across $4$ datasets after implementing attacks. 
Besides comparing with SC Trojan, BASS, and IHTG, we also benchmarked its performance against a clean, uncompromised SC system.
The attack experiments were conducted under both standard communication conditions (Signal-to-Noise Ratio [\text{SNR}] $= 25$\,dB) and noisy environments (\text{SNR} $= 5$\,dB), with the compression ratio fixed at $\tfrac{1}{4}$ and the poisoning rate $\gamma$ set to $20\%$.
Dataset partitioning followed a $4:1$ ratio, with $40$\,k samples designated for training and $10$\,k for evaluation across all sets.
We configured the multi-target attack result number as $T=4$, and $T$ samples are randomly selected in advance from the dataset as malicious targets.
Note that our method remains effective even when $T$ takes other settings.
Tabs. \ref{atk-total-30db} and \ref{tab:multi-ASR_30db} display the results for $25$\,dB communication conditions, and Tabs. \ref{atk-total-5db} and \ref{tab:multi-ASR_5db} provide the results for $5$\,dB.


\begin{table*}[t]\small
\caption{
\textbf{Comparison with the State-of-the-Art SC backdoors} on MNIST, F-MNIST, CIFAR-10, and ImageNet tasks under standard SNR $= 25$\,dB (ASR$\uparrow$: $\%$; PSNR*$\uparrow$, PSNR$^\circ$, $\Delta$PSNR$\downarrow$: dB). We mark the best results in bold across different tuning methods.
Please \textit{cf.} Sec. \ref{comparison result} for detailed explanations. 
}
\label{atk-total-30db}
\centering
\scriptsize{
\resizebox{\linewidth}{!}{
\setlength\tabcolsep{3.pt}
\renewcommand\arraystretch{1.1}
\begin{tabular}{r||cccc|cccc|cccc|cccc}
\hline\thickhline
\rowcolor{gray!20}
 & \multicolumn{4}{c|}{$\textbf{MNIST Dataset}$} & \multicolumn{4}{c|}{$\textbf{F-MNIST Dataset}$} & \multicolumn{4}{c|}{$\textbf{CIFAR-10 Dataset}$} & \multicolumn{4}{c}{$\textbf{ImageNet Dataset}$} \\
\rowcolor{gray!20}
\multirow{-2}{*}{\textbf{Methods}} & ASR & PSNR* & PSNR$^\circ$ & $\Delta$\text{PSNR} 
& ASR & PSNR* & PSNR$^\circ$ & $\Delta$\text{PSNR} 
& ASR & PSNR* & PSNR$^\circ$ & $\Delta$\text{PSNR} 
& ASR & PSNR* & PSNR$^\circ$ & $\Delta$\text{PSNR} \\
\hline\hline
\multicolumn{17}{l}{\textcolor{gray!60}{\textit{Attacking \textbf{JSCC} architecture}}} \\
Clean
& 0.00 & 36.47 & 36.47 & 0.00
& 0.00 & 33.82 & 33.82 & 0.00
& 0.00 & 31.15 & 31.15 & 0.00
& 0.00 & 34.93 & 34.93 & 0.00
\\
\hdashline
\rowcolor{gray!10} 
SC Trojan
& 85.76 & 33.33 & 36.47 & 3.14
& 91.81 & 29.60 & 33.82 & 4.22
& 86.68 & 25.78 & 31.15 & 5.37
& 89.88 & 32.08 & 34.93 & 2.85
\\
BASS
& 93.65 & 33.90 & 36.47 & 2.57
& 95.74 & 29.96 & 33.82 & 3.86
& 93.62 & 26.03 & 31.15 & 5.12
& 95.86 & 30.69 & 34.93 & 4.24
\\
\rowcolor{gray!10} 
IHTG
& 95.71 & 34.45 & 36.47 & 2.02
& 96.77 & 29.88 & 33.82 & 3.94
& 93.59 & 29.30 & 31.15 & 1.85
& 97.85 & 30.81 & 34.93 & 4.12
\\
\hline
\rowcolor[HTML]{D7F6FF}
\textbf{SemBugger} & \textbf{99.92} & \textbf{35.92} & {36.47} & \textbf{0.55} & \textbf{99.94} & \textbf{32.17} & {33.82} & \textbf{1.65} & \textbf{99.91} & \textbf{29.05} & {31.15} & \textbf{2.10} & \textbf{99.96} & \textbf{34.18} & {34.93} & \textbf{0.75} \\
\hline\hline
\multicolumn{17}{l}{\textcolor{gray!60}{\textit{Attacking \textbf{JSCC-f} architecture}}} \\
Clean
& 0.00 & 38.76 & 38.76 & 0.00
& 0.00 & 36.17 & 36.17 & 0.00
& 0.00 & 33.42 & 33.42 & 0.00
& 0.00 & 37.23 & 37.23 & 0.00
\\
\hdashline
\rowcolor{gray!10}
SC Trojan
& 88.75 & 32.87 & 38.76 & 5.89
& 92.80 & 32.71 & 36.17 & 3.46
& 89.67 & 29.41 & 33.42 & 4.01
& 91.87 & 34.26 & 37.23 & 2.97
\\
BASS
& 96.64 & 35.66 & 38.76 & 3.10
& 94.73 & 30.77 & 36.17 & 5.40
& 95.61 & 30.67 & 33.42 & 2.75
& 96.85 & 32.25 & 37.23 & 4.98
\\
\rowcolor{gray!10}
IHTG
& 94.70 & 35.76 & 38.76 & 3.00
& 93.76 & 33.70 & 36.17 & 2.47
& 93.58 & 29.07 & 33.42 & 4.35
& 95.84 & 35.43 & 37.23 & 1.80
\\
\hline
\rowcolor[HTML]{D7F6FF}
\textbf{SemBugger} & \textbf{99.93} & \textbf{37.66} & {38.76} & \textbf{1.10} & \textbf{99.95} & \textbf{35.32} & {36.17} & \textbf{0.85} & \textbf{99.92} & \textbf{31.22} & {33.42} & \textbf{2.20} & \textbf{99.97} & \textbf{35.28} & {37.23} & \textbf{1.95} \\
\hline\hline
\multicolumn{17}{l}{\textcolor{gray!60}{\textit{Attacking \textbf{JSCC-q} architecture}}} \\
Clean
& 0.00 & 38.53 & 38.53 & 0.00
& 0.00 & 35.82 & 35.82 & 0.00
& 0.00 & 33.14 & 33.14 & 0.00
& 0.00 & 36.91 & 36.91 & 0.00
\\
\hdashline
\rowcolor{gray!10}
SC Trojan
& 91.74 & 33.78 & 38.53 & 4.75
& 92.79 & 30.52 & 35.82 & 5.30
& 89.66 & 30.09 & 33.14 & 3.05
& 90.86 & 34.03 & 36.91 & 2.88
\\
BASS
& 97.63 & 35.54 & 38.53 & 2.99
& 95.72 & 31.32 & 35.82 & 4.50
& 97.60 & 27.99 & 33.14 & 5.15
& 96.84 & 33.51 & 36.91 & 3.40
\\
\rowcolor{gray!10}
IHTG
& 95.69 & 36.35 & 38.53 & 2.18
& 94.75 & 31.82 & 35.82 & 4.00
& 95.57 & 29.87 & 33.14 & 3.27
& 95.83 & 35.02 & 36.91 & 1.89
\\
\hline
\rowcolor[HTML]{D7F6FF}
\textbf{SemBugger} & \textbf{99.94} & \textbf{37.53} & {38.53} & \textbf{1.00} & \textbf{99.96} & \textbf{33.67} & {35.82} & \textbf{2.15} & \textbf{99.93} & \textbf{32.54} & {33.14} & \textbf{0.60} & \textbf{99.98} & \textbf{35.11} & {36.91} & \textbf{1.80} \\
\hline\hline
\multicolumn{17}{l}{\textcolor{gray!60}{\textit{Attacking \textbf{SCAN} architecture}}} \\
Clean
& 0.00 & 40.23 & 40.23 & 0.00
& 0.00 & 37.54 & 37.54 & 0.00
& 0.00 & 34.83 & 34.83 & 0.00
& 0.00 & 38.62 & 38.62 & 0.00
\\
\hdashline
\rowcolor{gray!10}
SC Trojan
& 84.73 & 37.52 & 40.23 & 2.71
& 90.78 & 31.64 & 37.54 & 5.90
& 85.65 & 30.39 & 34.83 & 4.44
& 87.85 & 34.95 & 38.62 & 3.67
\\
BASS
& 92.62 & 35.23 & 40.23 & 5.00
& 93.71 & 34.98 & 37.54 & 2.56
& 91.59 & 30.71 & 34.83 & 4.12
& 94.83 & 34.73 & 38.62 & 3.89
\\
\rowcolor{gray!10}
IHTG
& 91.68 & 36.43 & 40.23 & 3.80
& 93.74 & 35.52 & 37.54 & 2.02
& 92.56 & 32.88 & 34.83 & 1.95
& 93.82 & 34.42 & 38.62 & 4.20
\\
\hline
\rowcolor[HTML]{D7F6FF}
\textbf{SemBugger} & \textbf{99.95} & \textbf{38.23} & {40.23} & \textbf{2.00} & \textbf{99.97} & \textbf{37.04} & {37.54} & \textbf{0.50} & \textbf{99.94} & \textbf{33.53} & {34.83} & \textbf{1.30} & \textbf{99.99} & \textbf{36.87} & {38.62} & \textbf{1.75} \\
\hline\hline
\multicolumn{17}{l}{\textcolor{gray!60}{\textit{Attacking \textbf{SemCC} architecture}}} \\
Clean
& 0.00 & 39.20 & 39.20 & 0.00
& 0.00 & 39.34 & 39.34 & 0.00
& 0.00 & 39.82 & 39.82 & 0.00
& 0.00 & 39.62 & 39.62 & 0.00 \\
\hdashline
\rowcolor{gray!10}
SC Trojan
& 89.72 & 35.09 & 39.20 & 4.11
& 87.77 & 33.81 & 39.34 & 5.53
& 85.64 & 36.53 & 39.82 & 3.29
& 90.84 & 36.88 & 39.62 & 2.74 \\
BASS
& 92.61 & 36.17 & 39.20 & 3.03
& 91.70 & 34.21 & 39.34 & 5.13
& 93.58 & 35.46 & 39.82 & 4.36
& 93.82 & 36.97 & 39.62 & 2.65 \\
\rowcolor{gray!10}
IHTG
& 93.67 & 36.56 & 39.20 & 2.64
& 92.73 & 36.01 & 39.34 & 3.33
& 92.55 & 35.63 & 39.82 & 4.19
& 93.81 & 37.71 & 39.62 & 1.91 \\
\hline
\rowcolor[HTML]{D7F6FF}
\textbf{SemBugger} & \textbf{99.96} & \textbf{38.30} & {39.20} & \textbf{0.90} & \textbf{99.98} & \textbf{37.84} & {39.34} & \textbf{1.50} & \textbf{99.95} & \textbf{37.77} & {39.82} & \textbf{2.05} & \textbf{100.00} & \textbf{38.97} & {39.62} & \textbf{0.65} \\
\hline\thickhline
\end{tabular}}}
\end{table*}

\begin{table*}[t]
\small
\captionsetup{font=small}
\caption{\textbf{Target level ASR experimental results} on MNIST, F-MNIST, CIFAR-10, and ImageNet tasks under standard SNR $= 25$\,dB (ASR$\uparrow$: $\%$). Please \textit{cf.} Sec. \ref{comparison result} for detailed explanations.}
\centering
\label{tab:multi-ASR_30db}
\resizebox{\textwidth}{!}{%
\setlength\tabcolsep{3pt}
\renewcommand\arraystretch{1.1}
\begin{tabular}{r||cccc|cccc|cccc|cccc}
\hline \thickhline
\rowcolor{gray!20}
& \multicolumn{4}{c|}{\textbf{MNIST Dataset}} & \multicolumn{4}{c|}{\textbf{F-MNIST Dataset}} & \multicolumn{4}{c|}{\textbf{CIFAR-10 Dataset}} & \multicolumn{4}{c}{\textbf{ImageNet Dataset}} \\
\cline{2-17}
\rowcolor{gray!20}
\multirow{-2}{*}{\textbf{System}}  
& ASR$_1$ & ASR$_2$ & ASR$_3$ & ASR$_4$ 
& ASR$_1$ & ASR$_2$ & ASR$_3$ & ASR$_4$
& ASR$_1$ & ASR$_2$ & ASR$_3$ & ASR$_4$ 
& ASR$_1$ & ASR$_2$ & ASR$_3$ & ASR$_4$ \\
\hline\hline
JSCC     & 99.85 & 99.97 & 99.94 & 99.92 & 99.92 & 99.97 & 99.89 & 99.98 & 99.91 & 99.90 & 99.86 & 99.96 & 99.94 & 99.96 & 99.90 & 100.00 \\
\rowcolor{gray!10}
JSCC-f   & 99.91 & 99.94 & 99.76 & 100.00 & 99.93 & 99.94 & 99.92 & 100.00 & 99.88 & 99.87 & 99.80 & 99.82 & 99.89 & 99.94 & 99.90 & 100.00 \\
JSCC-q   & 99.90 & 99.92 & 99.84 & 99.88 & 99.94 & 99.92 & 99.90 & 99.92 & 99.87 & 99.95 & 99.88 & 99.93 & 99.95 & 99.97 & 99.89 & 99.96 \\
\rowcolor{gray!10}
SCAN     & 99.88 & 99.90 & 99.83 & 100.00 & 99.94 & 99.96 & 99.87 & 99.96 & 99.91 & 99.95 & 99.87 & 99.92 & 99.99 & 100.00 & 99.94 & 99.99 \\
SemCC    & 99.96 & 99.97 & 99.93 & 99.99 & 99.97 & 99.98 & 99.94 & 99.99 & 99.93 & 99.97 & 99.94 & 99.97 & 100.00 & 100.00 & 100.00 & 100.00 \\
\hline
\rowcolor[HTML]{D7F6FF}
\textbf{Average} & 99.90 & 99.94 & 99.86 & 99.96 & 99.94 & 99.95 & 99.90 & 99.97 & 99.90 & 99.93 & 99.87 & 99.92 & 99.95 & 99.97 & 99.93 & 99.99 \\
\hline \thickhline
\end{tabular}
}
\end{table*}

Concerning the test results under $25$\,dB, we get the following observations.
$1)$ The experimental results confirm that SemBugger achieves better attack performance while maintaining minimal impact on the original SC systems (Tab. \ref{atk-total-30db}). 
Concerning attack efficacy, SemBugger consistently attains near-perfect success rates (ASR $> 99.9\%$ across all datasets and SC architectures), wholly outperforming baseline backdoor attacks (SC Trojan, BASS, and IHTG), which typically depicts ASR between $85\%$ and $97\%$. 
On the other hand, regarding system fidelity preservation (\textit{i.e.}, attack stealthiness), SemBugger expresses the least PSNR degradation ($\Delta$PSNR), with average reductions of only $1.19$\,dB across all test cases,
where it is at least a $50\%$ reduction in distortion contrasted with baseline methods, which averages between $2.5$\,dB and $4.5$\,dB degradation. 
This is because we not only uphold the training paradigm for benign samples but also exploit a contrastive loss to separate their representations from those of the poisoned.
Especially noteworthy is its performance on ImageNet with the JSCC architecture, where it maintains a high PSNR$^\ast$ of $34.18$\,dB (compared to $34.93$\,dB for clean data), incurring only a $0.75$\,dB drop while achieving $99.96\%$ ASR. 
$2)$ The results manifest that SemBugger gains high ASRs across multiple target scenarios. 
As shown in Tab. \ref{tab:multi-ASR_30db}, across $4$ different datasets (MNIST, F-MNIST, CIFAR-10, and ImageNet) and $5$ SC architectures (JSCC, JSCC-f, JSCC-q, SCAN, and SemCC), SemBugger keeps an average ASR exceeding $99.0\%$ for all $4$ attack targets. Notably, it reaches an exceptional $99.96\%$ mean ASR on ImageNet set. 
Furthermore, the variation in ASR between different attack targets is minimal (maximum gap $= 0.15\%$), by which it indicates SemBugger's stable performance across diverse targets. 
This combination of high ASR and negligible quality deterioration is consistent across all $5$ architectures (JSCC, JSCC-f, JSCC-q, SCAN, and SemCC), which clearly proves SemBugger's advantage in both attack potency and stealthiness.


\begin{table*}[t]\small
\caption{
\textbf{Comparison with the State-of-the-Art SC backdoors} on MNIST, F-MNIST, CIFAR-10, and ImageNet tasks under constrained SNR $=5$\,dB (ASR: $\%$; PSNR*, PSNR$^\circ$, $\Delta$PSNR: dB).
We mark the best results in bold across different tuning methods.
Please \textit{cf.} Sec. \ref{comparison result} for detailed explanations.
}
\centering
\label{atk-total-5db}
\scriptsize
\resizebox{\linewidth}{!}{
\setlength\tabcolsep{3pt}
\renewcommand\arraystretch{1.1}
\begin{tabular}{r||cccc|cccc|cccc|cccc}
\hline\thickhline
\rowcolor{gray!20}
 & \multicolumn{4}{c|}{\textbf{MNIST Dataset}}
 & \multicolumn{4}{c|}{\textbf{F-MNIST Dataset}}
 & \multicolumn{4}{c|}{\textbf{CIFAR-10 Dataset}}
 & \multicolumn{4}{c}{\textbf{ImageNet Dataset}}\\
\rowcolor{gray!20}
\multirow{-2}{*}{\textbf{Methods}}
& ASR & PSNR* & PSNR$^\circ$ & $\Delta$PSNR
& ASR & PSNR* & PSNR$^\circ$ & $\Delta$PSNR
& ASR & PSNR* & PSNR$^\circ$ & $\Delta$PSNR
& ASR & PSNR* & PSNR$^\circ$ & $\Delta$PSNR \\
\hline\hline
\multicolumn{17}{l}{\textcolor{gray!60}{\textit{Attacking \textbf{JSCC} architecture}}} \\
Clean       & 0.00 & 21.47 & 21.47 & 0.00 & 0.00 & 19.82 & 19.82 & 0.00 & 0.00 & 14.15 & 14.15 & 0.00 & 0.00 & 16.93 & 16.93 & 0.00 \\
\hdashline
\rowcolor{gray!10} SC Trojan   & 76.39 & 16.32 & 21.47 & 5.15 & 79.36 & 14.02 & 19.82 & 5.80 & 71.46 & 7.61  & 14.15 & 6.54 & 79.29 & 11.41 & 16.93 & 5.52 \\
BASS        & 89.37 & 17.85 & 21.47 & 3.62 & 87.68 & 16.23 & 19.82 & 3.59 & 86.87 & 10.02 & 14.15 & 4.13 & 90.27 & 13.43 & 16.93 & 3.50 \\
\rowcolor{gray!10} IHTG        & 90.58 & 18.35 & 21.47 & 3.12 & 89.28 & 16.57 & 19.82 & 3.25 & 89.91 & 10.56 & 14.15 & 3.59 & 91.58 & 13.62 & 16.93 & 3.31 \\
\hline
\rowcolor[HTML]{D7F6FF}SemBugger & \textbf{97.47} & \textbf{19.79} & 21.47 & \textbf{1.68} & \textbf{95.65} & \textbf{18.23} & 19.82 & \textbf{1.59} & \textbf{94.08} & \textbf{12.65} & 14.15 & \textbf{1.50} & \textbf{96.29} & \textbf{15.34} & 16.93 & \textbf{1.59} \\
\hline\hline
\multicolumn{17}{l}{\textcolor{gray!60}{\textit{Attacking \textbf{JSCC-f} architecture}}} \\
Clean       & 0.00 & 23.76 & 23.76 & 0.00 & 0.00 & 21.17 & 21.17 & 0.00 & 0.00 & 16.42 & 16.42 & 0.00 & 0.00 & 19.23 & 19.23 & 0.00 \\
\hdashline
\rowcolor{gray!10} SC Trojan   & 71.97 & 17.92 & 23.76 & 5.84 & 83.75 & 16.21 & 21.17 & 4.96 & 75.45 & 9.81  & 16.42 & 6.61 & 80.54 & 13.48 & 19.23 & 5.75 \\
BASS        & 89.21 & 19.63 & 23.76 & 4.13 & 90.61 & 17.32 & 21.17 & 3.85 & 86.62 & 12.31 & 16.42 & 4.11 & 91.48 & 15.42 & 19.23 & 3.81 \\
\rowcolor{gray!10} IHTG        & 90.58 & 19.87 & 23.76 & 3.89 & 86.88 & 17.64 & 21.17 & 3.53 & 90.11 & 12.89 & 16.42 & 3.53 & 88.29 & 15.68 & 19.23 & 3.55 \\
\hline
\rowcolor[HTML]{D7F6FF}SemBugger & \textbf{96.95} & \textbf{21.63} & 23.76 & \textbf{2.13} & \textbf{94.51} & \textbf{19.66} & 21.17 & \textbf{1.51} & \textbf{96.16} & \textbf{15.03} & 16.42 & \textbf{1.39} & \textbf{95.12} & \textbf{17.62} & 19.23 & \textbf{1.61} \\
\hline\hline
\multicolumn{17}{l}{\textcolor{gray!60}{\textit{Attacking \textbf{JSCC-q} architecture}}} \\
Clean       & 0.00 & 23.53 & 23.53 & 0.00 & 0.00 & 20.82 & 20.82 & 0.00 & 0.00 & 16.14 & 16.14 & 0.00 & 0.00 & 20.91 & 20.91 & 0.00 \\
\hdashline
\rowcolor{gray!10} SC Trojan   & 81.75 & 17.51 & 23.53 & 6.02 & 80.92 & 14.21 & 20.82 & 6.61 & 72.32 & 10.11 & 16.14 & 6.03 & 77.33 & 12.82 & 20.91 & 8.09 \\
BASS        & 93.05 & 19.41 & 23.53 & 4.12 & 86.91 & 17.21 & 20.82 & 3.61 & 91.27 & 11.59 & 16.14 & 4.55 & 91.40 & 17.39 & 20.91 & 3.52 \\
\rowcolor{gray!10} IHTG        & 90.02 & 19.78 & 23.53 & 3.75 & 91.41 & 17.57 & 20.82 & 3.25 & 88.45 & 11.91 & 16.14 & 4.23 & 90.85 & 16.91 & 20.91 & 4.00 \\
\hline
\rowcolor[HTML]{D7F6FF}SemBugger & \textbf{94.41} & \textbf{21.71} & 23.53 & \textbf{1.82} & \textbf{97.59} & \textbf{19.49} & 20.82 & \textbf{1.33} & \textbf{96.04} & \textbf{14.79} & 16.14 & \textbf{1.35} & \textbf{95.80} & \textbf{19.19} & 20.91 & \textbf{1.72} \\
\hline\hline
\multicolumn{17}{l}{\textcolor{gray!60}{\textit{Attacking \textbf{SCAN} architecture}}} \\
Clean       & 0.00 & 23.23 & 23.23 & 0.00 & 0.00 & 20.54 & 20.54 & 0.00 & 0.00 & 15.83 & 15.83 & 0.00 & 0.00 & 19.62 & 19.62 & 0.00 \\
\hdashline
\rowcolor{gray!10} SC Trojan   & 72.19 & 17.31 & 23.23 & 5.92 & 73.67 & 13.41 & 20.54 & 7.13 & 76.42 & 9.71  & 15.83 & 6.12 & 72.97 & 11.61 & 19.62 & 8.01 \\
BASS        & 85.03 & 19.38 & 23.23 & 3.85 & 88.49 & 16.53 & 20.54 & 4.01 & 82.65 & 11.79 & 15.83 & 4.04 & 90.66 & 15.61 & 19.62 & 4.01 \\
\rowcolor{gray!10} IHTG        & 83.66 & 19.53 & 23.23 & 3.70 & 89.97 & 16.81 & 20.54 & 3.73 & 85.72 & 12.01 & 15.83 & 3.82 & 88.33 & 15.83 & 19.62 & 3.79 \\
\hline
\rowcolor[HTML]{D7F6FF}SemBugger & \textbf{95.28} & \textbf{21.52} & 23.23 & \textbf{1.71} & \textbf{97.81} & \textbf{19.03} & 20.54 & \textbf{1.51} & \textbf{94.44} & \textbf{14.51} & 15.83 & \textbf{1.32} & \textbf{96.70} & \textbf{18.17} & 19.62 & \textbf{1.45} \\
\hline\hline
\multicolumn{17}{l}{\textcolor{gray!60}{\textit{Attacking \textbf{SemCC} architecture}}} \\
Clean       & 0.00 & 22.20 & 22.20 & 0.00 & 0.00 & 22.34 & 22.34 & 0.00 & 0.00 & 22.82 & 22.82 & 0.00 & 0.00 & 22.62 & 22.62 & 0.00 \\
\hdashline
\rowcolor{gray!10} SC Trojan   & 80.55 & 16.35 & 22.20 & 5.85 & 74.32 & 16.55 & 22.34 & 5.79 & 67.75 & 14.21 & 22.82 & 8.61 & 77.28 & 16.71 & 22.62 & 5.91 \\
BASS        & 86.43 & 18.76 & 22.20 & 3.44 & 82.74 & 18.81 & 22.34 & 3.53 & 89.21 & 19.39 & 22.82 & 3.43 & 86.57 & 19.18 & 22.62 & 3.44 \\
\rowcolor{gray!10} IHTG        & 89.11 & 18.94 & 22.20 & 3.26 & 84.84 & 18.92 & 22.34 & 3.42 & 89.34 & 18.71 & 22.82 & 4.11 & 87.08 & 18.41 & 22.62 & 4.21 \\
\hline
\rowcolor[HTML]{D7F6FF}SemBugger & \textbf{94.74} & \textbf{20.76} & 22.20 & \textbf{1.44} & \textbf{96.84} & \textbf{20.88} & 22.34 & \textbf{1.46} & \textbf{97.19} & \textbf{21.39} & 22.82 & \textbf{1.43} & \textbf{98.00} & \textbf{21.18} & 22.62 & \textbf{1.44} \\
\hline\thickhline
\end{tabular}}
\end{table*}

\begin{table*}[t]
\small
\captionsetup{font=small}
\caption{\textbf{Target level ASR experimental results} on MNIST, F-MNIST, CIFAR-10, and ImageNet tasks under constrained SNR $= 5$\,dB (ASR $\uparrow$: $\%$). Please $\textit{cf.}$ Sec. \ref{comparison result} for detailed explanations.}
\centering
\label{tab:multi-ASR_5db}
\resizebox{\textwidth}{!}{%
\setlength\tabcolsep{3pt}
\renewcommand\arraystretch{1.1}
\begin{tabular}{r||cccc|cccc|cccc|cccc}
\hline \thickhline
\rowcolor{gray!20}
& \multicolumn{4}{c|}{\textbf{MNIST Dataset}} & \multicolumn{4}{c|}{\textbf{F-MNIST Dataset}} & \multicolumn{4}{c|}{\textbf{CIFAR-10 Dataset}} & \multicolumn{4}{c}{\textbf{ImageNet Dataset}} \\
\cline{2-17}
\rowcolor{gray!20}
\multirow{-2}{*}{\textbf{System}}  
& ASR$_1$ & ASR$_2$ & ASR$_3$ & ASR$_4$ 
& ASR$_1$ & ASR$_2$ & ASR$_3$ & ASR$_4$
& ASR$_1$ & ASR$_2$ & ASR$_3$ & ASR$_4$
& ASR$_1$ & ASR$_2$ & ASR$_3$ & ASR$_4$\\
\hline\hline
JSCC     & 97.84 & 96.52 & 97.13 & 98.41 & 95.26 & 95.79 & 95.88 & 96.47 & 93.77 & 94.58 & 94.11 & 93.92 & 95.68 & 96.03 & 96.36 & 97.21 \\
\rowcolor{gray!10}
JSCC-f   & 96.32 & 97.21 & 96.86 & 97.38 & 93.65 & 94.48 & 95.24 & 94.59 & 95.67 & 96.36 & 96.09 & 96.61 & 94.25 & 95.18 & 95.76 & 95.42 \\
JSCC-q   & 93.89 & 94.71 & 95.14 & 94.03 & 97.52 & 97.91 & 97.33 & 97.69 & 95.93 & 96.28 & 96.17 & 95.86 & 95.27 & 95.96 & 95.65 & 96.66 \\
\rowcolor{gray!10}
SCAN     & 94.51 & 95.08 & 95.29 & 96.34 & 96.85 & 97.46 & 97.77 & 97.03 & 93.81 & 94.13 & 94.96 & 94.37 & 96.12 & 96.57 & 96.65 & 97.29 \\
SemCC    & 93.83 & 94.64 & 94.91 & 95.57 & 96.35 & 97.12 & 96.68 & 97.18 & 96.86 & 97.61 & 96.48 & 97.63 & 97.71 & 97.89 & 98.07 & 98.28 \\
\hline
\rowcolor[HTML]{D7F6FF}
\textbf{Average} & 95.28 & 95.63 & 95.87 & 96.35 & 95.93 & 96.55 & 96.58 & 96.59 & 95.21 & 95.79 & 95.56 & 95.68 & 95.81 & 96.33 & 96.50 & 96.97 \\
\hline \thickhline
\end{tabular}}
\end{table*}

Tabs. ~\ref{atk-total-5db} and \ref{tab:multi-ASR_5db} report quantitative results under a constrained SNR of $5$\,dB, comparing SemBugger with $3$ attack baselines across JSCC, JSCC-f, JSCC-q, SCAN, and SemCC and $4$ benchmark MNIST, F-MNIST, CIFAR-10, and ImageNet sets. 
We have some key analysis.
$1)$ SemBugger consistently reaches the best ASR values, ranging from $94.08$\% to $98.00$\%, while retaining lower PSNR decline than baselines. 
Specifically, the average $\Delta$PSNR caused by SemBugger remains between $1.32$\,dB and $2.13$\,dB across all test cases, whereby it is substantially less than SC Trojan (maxima $=8.61$\,dB), BASS (maxima $=4.55$\,dB), and IHTG (maxima $=4.23$\,dB). 
For instance, on ImageNet, SemBugger reduces PSNR by only $1.59$\,dB on JSCC and $1.44$\,dB on SemCC, while realizing ASR of $96.29$\% and $98.00$\%. 
Similar trends are observed through other datasets, with PSNR* values invariably closer to the clean baseline. 
While both SemBugger and the baselines express diminished ASR and increased $\Delta$PSNR under poorer communication conditions, their overall efficacy stays within acceptable bounds, with SemBugger demonstrating particularly stable behavior.
$2)$ Tab.~\ref{tab:multi-ASR_5db} summarizes the target-level ASR across $4$ datasets and $5$ SC architectures under an SNR of $5$\,dB. Overall, all systems achieve high target-specific ASR values, with most results passing $93$\% over all targets and datasets. Notably, SemCC has the highest per-target ASR on ImageNet, which reaches up to $98.28$\%. 
Congruent tendencies are gotten for other architectures, revealing that stable multi-target attack efficacy is maintained under tough communication conditions. 

\begin{figure}[t]
    \centering
    \includegraphics[width=0.91\linewidth, keepaspectratio]{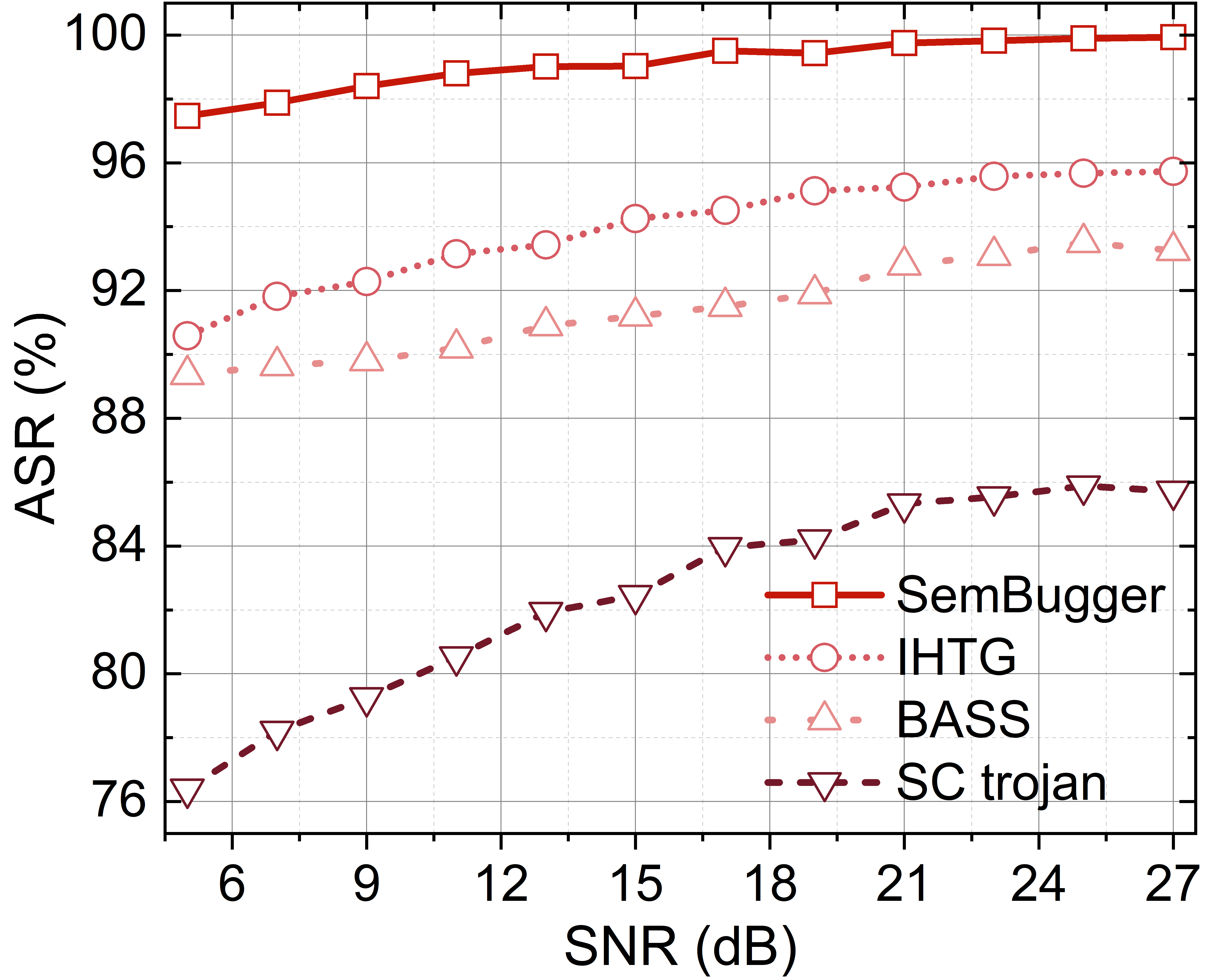}
    \caption{\textbf{Test results of ASRs across SNRs}  (ASR: $\%$; SNR: dB). Please \textit{cf.} Sec. \ref{comparison result} for detailed explanations.} 
    \label{fig:SNR-ASR}
\end{figure}

\begin{figure}[t]
    \centering
    \includegraphics[width=0.91\linewidth, keepaspectratio]{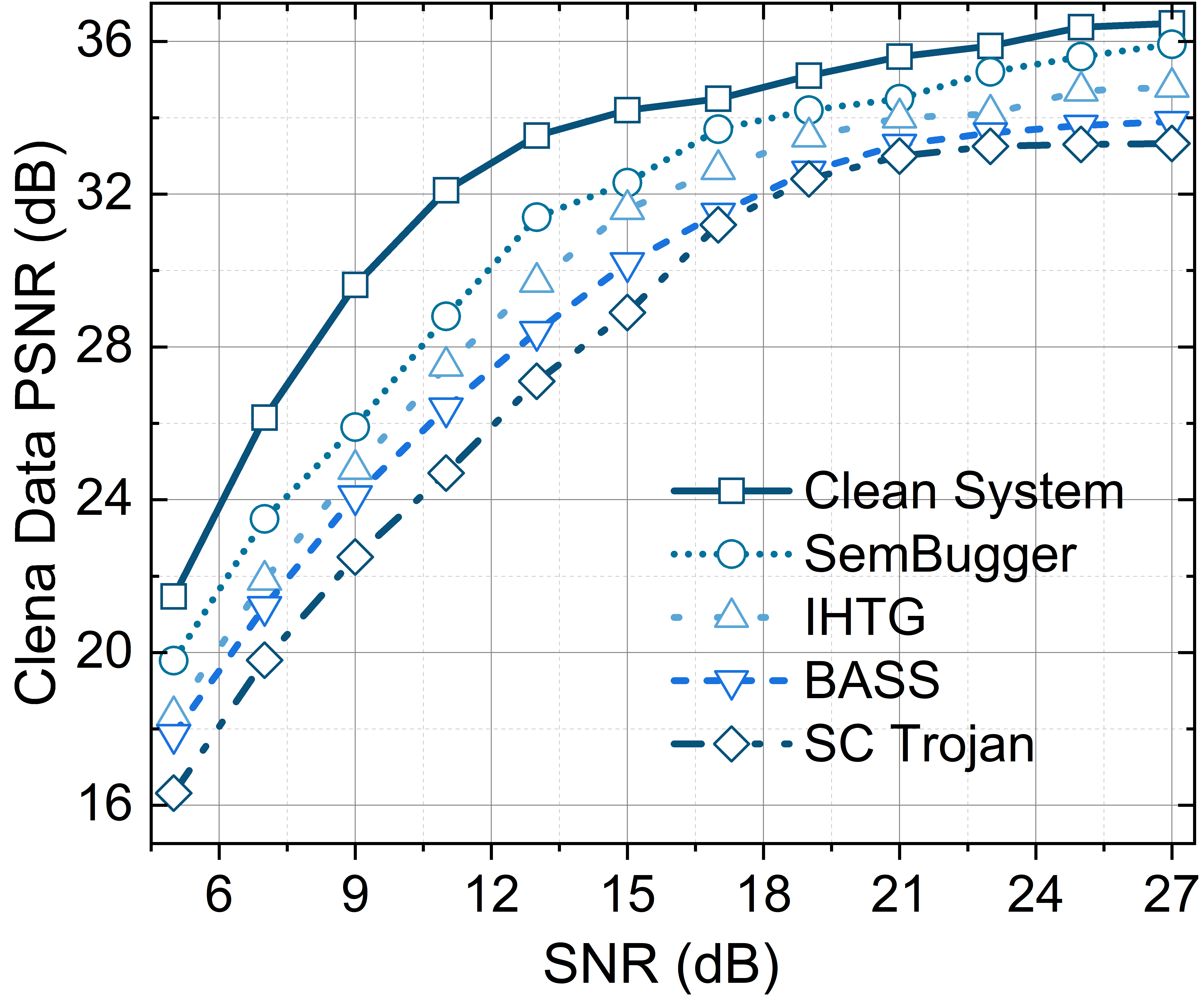}
    \caption{\textbf{Test results of clean data PSNRs across SNRs}  (SNR, PSNR: dB). Please \textit{cf.} Sec. \ref{comparison result} for detailed explanations.} 
    \label{fig:SNR-delta-PSNR}
\end{figure}

\vspace{0.8ex}
\noindent \textbf{Signal-to-Noise Ratio}. 
The impact of SNR under various communication conditions was further inspected. 
We compared the ASR and $\Delta$PSNR across various attacks. 
The experiments were executed using the JSCC architecture and the MNIST set. 
The results are presented in Figs. \ref{fig:SNR-ASR} and \ref{fig:SNR-delta-PSNR}.

As illustrated in Fig. \ref{fig:SNR-ASR} for attack efficacy, the ASR increases with higher SNR values for all methods, indicating that improved communication conditions lead to more effective attacks. 
SemBugger persistently outperforms the other methods, having the highest ASR at each SNR level. At an SNR $5$\,dB, SemBugger already has an ASR $=97.47\%$, and this value rises steadily. 
Conversely, IHTG, BASS, and SC Trojan show relatively lower ASRs, with their performance improving as the SNR increases but not reaching the levels of SemBugger. 
The results statistically validate SemBugger's ASR advantage under diverse channel conditions, particularly noise-resilient performance with $<3\%$ efficiency deviation through $15-35$\,dB SNR interval.

In Fig. \ref{fig:SNR-delta-PSNR}, we contrasted our SemBugger with IHTG, BASS, and SC Trojan in $\Delta$PSNR.
As the SNR grows, $\Delta$PSNR for all methods improves, and it manifests that the transmission efficiency of clean data becomes less affected by the attacks under better communication conditions. 
SemBugger sustains the lowest $\Delta$PSNR traversing all SNR levels. 
At SNR $=5$\,dB, SemBugger gets a $\Delta$PSNR $=19.79$\,dB, and this value increases to $35.92$\,dB at SNR $=27$\,dB. 
Differentially, the other methods present relatively larger $\Delta$PSNR values, whereby this implies more severe deterioration of system efficacy, especially at lower SNR settings. 

\begin{figure}[t]
    \centering

    \adjustbox{valign=c}{%
        \parbox{0.07\textwidth}{\centering\small Source\\Data}}%
    \hspace{0.005\textwidth}
    \adjustbox{valign=c}{\includegraphics[width=0.07\textwidth]{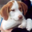}}\hspace{0.003\textwidth}
    \adjustbox{valign=c}{\includegraphics[width=0.07\textwidth]{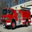}}\hspace{0.003\textwidth}
    \adjustbox{valign=c}{\includegraphics[width=0.07\textwidth]{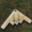}}\hspace{0.003\textwidth}
    \adjustbox{valign=c}{\includegraphics[width=0.07\textwidth]{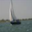}}\hspace{0.003\textwidth}
    \adjustbox{valign=c}{\includegraphics[width=0.07\textwidth]{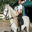}}\\[0.5em]

    \noindent\rule{0.48\textwidth}{0.5pt}\\[0.5em]

    \adjustbox{valign=c}{%
        \parbox{0.07\textwidth}{\centering\small \textbf{Our}\\\textbf{Method}}}%
    \hspace{0.005\textwidth}
    \adjustbox{valign=c}{\includegraphics[width=0.07\textwidth]{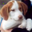}}\hspace{0.003\textwidth}
    \adjustbox{valign=c}{\includegraphics[width=0.07\textwidth]{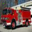}}\hspace{0.003\textwidth}
    \adjustbox{valign=c}{\includegraphics[width=0.07\textwidth]{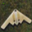}}\hspace{0.003\textwidth}
    \adjustbox{valign=c}{\includegraphics[width=0.07\textwidth]{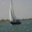}}\hspace{0.003\textwidth}
    \adjustbox{valign=c}{\includegraphics[width=0.07\textwidth]{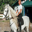}}\\[0.5em]

    \adjustbox{valign=c}{%
        \parbox{0.07\textwidth}{\centering\small SC Trojan}}%
    \hspace{0.005\textwidth}
    \adjustbox{valign=c}{\includegraphics[width=0.07\textwidth]{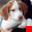}}\hspace{0.003\textwidth}
    \adjustbox{valign=c}{\includegraphics[width=0.07\textwidth]{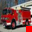}}\hspace{0.003\textwidth}
    \adjustbox{valign=c}{\includegraphics[width=0.07\textwidth]{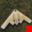}}\hspace{0.003\textwidth}
    \adjustbox{valign=c}{\includegraphics[width=0.07\textwidth]{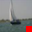}}\hspace{0.003\textwidth}
    \adjustbox{valign=c}{\includegraphics[width=0.07\textwidth]{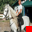}}\\[0.5em]

    \adjustbox{valign=c}{%
        \parbox{0.07\textwidth}{\centering\small BASS}}%
    \hspace{0.005\textwidth}
    \adjustbox{valign=c}{\includegraphics[width=0.07\textwidth]{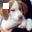}}\hspace{0.003\textwidth}
    \adjustbox{valign=c}{\includegraphics[width=0.07\textwidth]{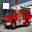}}\hspace{0.003\textwidth}
    \adjustbox{valign=c}{\includegraphics[width=0.07\textwidth]{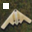}}\hspace{0.003\textwidth}
    \adjustbox{valign=c}{\includegraphics[width=0.07\textwidth]{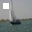}}\hspace{0.003\textwidth}
    \adjustbox{valign=c}{\includegraphics[width=0.07\textwidth]{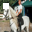}}\\[0.5em]

    \adjustbox{valign=c}{%
        \parbox{0.07\textwidth}{\centering\small IHTG}}%
    \hspace{0.005\textwidth}
    \adjustbox{valign=c}{\includegraphics[width=0.07\textwidth]{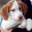}}\hspace{0.003\textwidth}
    \adjustbox{valign=c}{\includegraphics[width=0.07\textwidth]{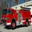}}\hspace{0.003\textwidth}
    \adjustbox{valign=c}{\includegraphics[width=0.07\textwidth]{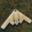}}\hspace{0.003\textwidth}
    \adjustbox{valign=c}{\includegraphics[width=0.07\textwidth]{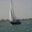}}\hspace{0.003\textwidth}
    \adjustbox{valign=c}{\includegraphics[width=0.07\textwidth]{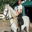}}

    \caption{\textbf{Illustration of attack (\textit{i.e.}, poisoned) data} generated across backdoors: our SemBugger, SC Trojan, BASS, and baseline source data.
    $1)$ The poisonous data via our SemBugger exhibits negligible perceptual divergence from the source data. $2)$ The attack data generated by SC Trojan and BASS expose conspicuous visual artifacts. $3)$ SemBugger matches the concealment capability of SOTA attacks like IHTG.}
    \label{sample-illu}
\end{figure}

\vspace{0.8ex}
\noindent \textbf{Sample Visualization}. 
To more effectively highlight the imperceptibility of our backdoor triggers (mentioned in Sec.~\ref{design_require}), we performed a comparison between adversarial examples generated by our SemBugger (w/ triggers of maximum level intensity) and baselines.
Fig. \ref{sample-illu} offers a visual exposition of poisoned samples from CIFAR-10 dataset and reflects the disparities in trigger stealth features. The attack was implemented under JSCC (w/ $25$\,dB channel), and the visualization results are showcased in Fig. \ref{sample-illu}.

From the figure, the samples validate that, for visual modality, the triggers from our SemBugger retain a markedly better semantic congruence with the source content. 
Subjectively, the implanted trigger pattern blends almost imperceptibly into the data scene, while quantitatively SSIM reaches $95.25\%$ for SemBugger, which is above BASS ($80.52\%$) and SC Trojan ($78.63\%$), and approximates IHTG ($95.98\%$). 
These findings substantiates that SemBugger has good perceptual camouflage without diminishing attack efficacy while attaining performance analogous to SOTA covert backdoor attacks (\textit{i.e.}, IHTG) when compared with prevailing SC backdoors.

\vspace{0.8ex}
\noindent \textbf{Regular Task Impact}. 
We evaluate the impact of the attack on system nominal performance under poor channel conditions (SNR $= 5$\,dB). We appended a classifier (RegNetY-16GF network) to the system and launched the attack against this end-to-end pipeline. We then measured the classification efficacy of the clean system and the attacked system, assessing how the attack affects overall system behavior in a degraded transmission setting. 
The evaluation was conducted on the MNIST set, with the corresponding results shown in Tab. \ref{atk_task_influence}.

\begin{table}[t]\small
\caption{
\textbf{Regular Task classification efficacy after attacks} on MNIST tasks under poor SNR $= 5$\,dB ($\%$ $\uparrow$). Please \textit{cf.} Sec. \ref{comparison result} for detailed explanations.}

\centering
\label{atk_task_influence}
\scriptsize
\resizebox{\linewidth}{!}{
\setlength\tabcolsep{3pt}
\renewcommand\arraystretch{1.1}
\begin{tabular}{r||ccccc}
\hline\thickhline
\rowcolor{gray!20}
 & \multicolumn{5}{c}{\textbf{Test Systems}}\\
\rowcolor{gray!20}
\multirow{-2}{*}{\textbf{Condition}}
& JSCC & JSCC-f & JSCC-q & SCAN & SemCC \\
\hline\hline
Clean & 93.28 & 93.43 & 92.87 & 91.95 & 94.76 \\
\hline
\rowcolor{gray!10}
Attack & 93.02 & 93.18 & 92.62 & 91.68 & 94.05 \\
\hline\thickhline
\end{tabular}}
\end{table}

It can be observed that all evaluated systems maintain nearly identical classification efficiency before and after the attack. For Instance, JSCC decreases solely from $93.28\%$ to $93.02\%$ ($-0.26$ percentage points), JSCC-f from $93.43\%$ to $93.18\%$ ($-0.25$), JSCC-q from $92.87\%$ to $92.62\%$ ($-0.25$), SCAN from $91.95\%$ to $91.68\%$ ($-0.27$), and SemCC from $94.76\%$ to $94.05\%$ ($-0.71$). Overall, the performance degradation induced by the attack is consistently below $1\%$, implying that the downstream task capability under clean inputs is largely unaffected. This is because our attack does not alter the semantic representations transmitted for clean samples. Instead, it induces abnormal behavior only under specific trigger conditions. Consequently, in standard (non-triggered) evaluations, the semantic information received by the downstream classifier remains stable, and the classification accuracy exhibits negligible degradation.

\vspace{0.8ex}
\noindent \textbf{Spatial Processing Influence}. 
We further study the impact of input spatial cropping on attack performance. Using spatial cropping, we randomly crop portions of the original trigger and embed them into clean samples to assess how cropping degree affects attack efficacy. Experiments are conducted under $25$\,dB SNR using the JSCC on MNIST and Fashion-MNIST datasets. Four cropping ratios (\textit{i.e.}, $1/2$, $1/4$, $1/8$, and $1/16$) of the full trigger are tested to analyze the effect of cropping. To evaluate stealthiness, we adopt Learned Perceptual Image Patch Similarity (LPIPS) to measure perceptual disparity between attacked and original samples under multiple cropping conditions.
Empirical results are depicted in Tab.~\ref{cropping_test}.

\begin{table}[t]\small
\caption{
\textbf{Attack efficacy (ASR $\uparrow$) and stealthiness (LPIPS $\downarrow$)} on MNIST tasks under standard SNR $= 25$\,dB. Please \textit{cf.} Sec. \ref{comparison result} for detailed explanations. 
}
\centering
\label{cropping_test}
\scriptsize
\resizebox{0.8\linewidth}{!}{
\setlength\tabcolsep{3pt}
\renewcommand\arraystretch{1.1}
\begin{tabular}{r||cccc}
\hline\thickhline
\rowcolor{gray!20}
 & \multicolumn{4}{c}{\textbf{Cropping Ratios}}\\
\rowcolor{gray!20}
\multirow{-2}{*}{\textbf{Metric}}
& 1/2 & 1/4 & 1/8 & 1/16 \\
\hline\hline
ASR (\%) & 99.54 & 98.13 & 96.48 & 91.10 \\
\hline
\rowcolor{gray!10} LPIPS & 0.105 & 0.089 & 0.062 & 0.041 \\
\hline\thickhline
\end{tabular}}
\end{table}

In terms of attack efficacy, the proposed attack consistently achieves high ASR even in the presence of cropping. Concretely, ASR reaches $99.54\%$ and $98.13\%$ for cropping ratios of $1/2$ and $1/4$, respectively, and remains as high as $96.48\%$ and $91.10\%$ when the cropping ratio is increased to $1/8$ and $1/16$, whereby it proves strong robustness against cropping perturbations. 
Meanwhile, the LPIPS values remain low across all settings and decrease progressively from $0.105$ to $0.041$ as the cropping ratio increases, which shows that the attacked poisonous data are perceptually close to the clean ones and thus highly stealthy. 
Overall, empirical results suggest that the proposed attack remains highly effective and preserves strong imperceptibility, even in the presence of common preprocessing operations like data cropping.


\subsection{Ablation Study}
\label{ablation study}
We further ran ablation experiments to inspect various factors influencing our experimental results, with a focus on $1)$ compression rate $CR$ and $2)$ poisoning rate $\gamma$. The tests are mainly performed using the JSCC architecture.

\vspace{0.8ex}
\noindent \textbf{Compression Rate}. 
We examined ASR for $4$ attack target conditions by configuring multiple $CR$, while also quantitatively appraising the performance deterioration ($\Delta$PSNR) caused by the attack on benign data transmission quality.
Experiments were carried out employing the JSCC-f SC system under $25$\,dB SNR constraint.
Tab. ~\ref{tab:compression rate} presents the results across varying $CR$ for MNIST, F-MNIST, CIFAR-10, and ImageNet datasets.

\begin{table*}[t]
\small
\captionsetup{font=small}
\caption{\textbf{Ablation study for compression rates $CR$} on MNIST, F-MNIST, CIFAR-10, and ImageNet tasks under constrained SNR $= 25$\,dB (ASR$\uparrow$: $\%$, $\Delta$PSNR$\uparrow$: dB). Please \textit{cf.} Sec. \ref{ablation study} for detailed explanations.}
\centering
\label{tab:compression rate}
\resizebox{\textwidth}{!}{%
\setlength\tabcolsep{3pt}
\renewcommand\arraystretch{1.1}
\begin{tabular}{r||cccc|c|cccc|c|cccc|c|cccc|c}
\hline \thickhline
\rowcolor{gray!20}
& \multicolumn{5}{c|}{\textbf{MNIST Dataset}} & \multicolumn{5}{c|}{\textbf{F-MNIST Dataset}} & \multicolumn{5}{c|}{\textbf{CIFAR-10 Dataset}} & \multicolumn{5}{c}{\textbf{ImageNet Dataset}} \\
\cline{2-21}
\rowcolor{gray!20}
\multirow{-2}{*}{\textbf{$CR$}}  
& ASR$_1$ & ASR$_2$ & ASR$_3$ & ASR$_4$ & $\Delta$PSNR 
& ASR$_1$ & ASR$_2$ & ASR$_3$ & ASR$_4$ & $\Delta$PSNR 
& ASR$_1$ & ASR$_2$ & ASR$_3$ & ASR$_4$ & $\Delta$PSNR 
& ASR$_1$ & ASR$_2$ & ASR$_3$ & ASR$_4$ & $\Delta$PSNR \\
\hline\hline
1$/$12 & 92.35 & 93.42 & 91.88 & 92.17 & 1.17 & 91.47 & 93.15 & 90.82 & 92.64 & 2.31 & 88.23 & 89.71 & 87.95 & 90.12 & 2.39 & 89.47 & 90.33 & 88.92 & 91.25 & 2.03 \\
\rowcolor{gray!10}
1$/$4  & 99.91 & 99.94 & 99.76 & 100.00 & 1.10 & 99.93 & 99.94 & 99.92 & 100.00 & 0.85 & 99.88 & 99.87 & 99.80 & 99.82 & 2.20 & 99.89 & 99.94 & 99.90 & 100.00 & 1.95 \\
1$/$3  & 99.97 & 99.99 & 99.93 & 100.00 & 0.87 & 99.98 & 99.99 & 99.97 & 100.00 & 0.62 & 99.95 & 99.96 & 99.92 & 99.95 & 1.64 & 99.97 & 99.99 & 99.96 & 100.00 & 0.52 \\
\rowcolor{gray!10} 
2$/$5  & 99.99 & 100.00 & 99.97 & 100.00 & 0.28 & 99.99 & 100.00 & 99.98 & 100.00 & 0.24 & 99.97 & 99.98 & 99.95 & 99.97 & 1.13 & 99.99 & 100.00 & 99.98 & 100.00 & 0.41 \\
\hline
\rowcolor[HTML]{D7F6FF}
\textbf{Avg} & 98.63 & 98.86 & 98.49 & 98.83 & 0.85 & 98.67 & 98.99 & 98.53 & 98.93 & 1.01 & 97.79 & 98.29 & 97.90 & 98.15 & 1.84 & 97.85 & 98.25 & 97.94 & 98.45 & 1.22 \\
\hline \thickhline
\end{tabular}}
\end{table*}

With the results, $1)$ the multi-target attack reaches consistently high success rates (ASR\textsubscript{1}-ASR\textsubscript{4}) traversing all evaluated $CR$, and maintains $>88\%$ effectiveness even at the most extreme compression with $CR=1/12$.
$2)$ Performance improves monotonically with increasing bandwidth allocation, reaching stable success rates ($\geq99.76\%$) for $CR\geq1/4$ across all datasets and target results.
$3)$ The attack elicits minimal distortion to regular transmissions capability, with $\Delta$PSNR values constrained below $2.39$\,dB in all configurations.
Our evaluations reveal that SemBugger has consistent adversarial effectiveness and keeps stealthiness across regular task $CR$.
This stems from the joint multi-effect poisoning-training of our attack with the SC system, by which it forces the shared SC knowledge to sustain high ASRs without sacrificing system's benign data transmission functionality.

\begin{figure}[t]
    \centering
    \includegraphics[width=0.95\linewidth, keepaspectratio]{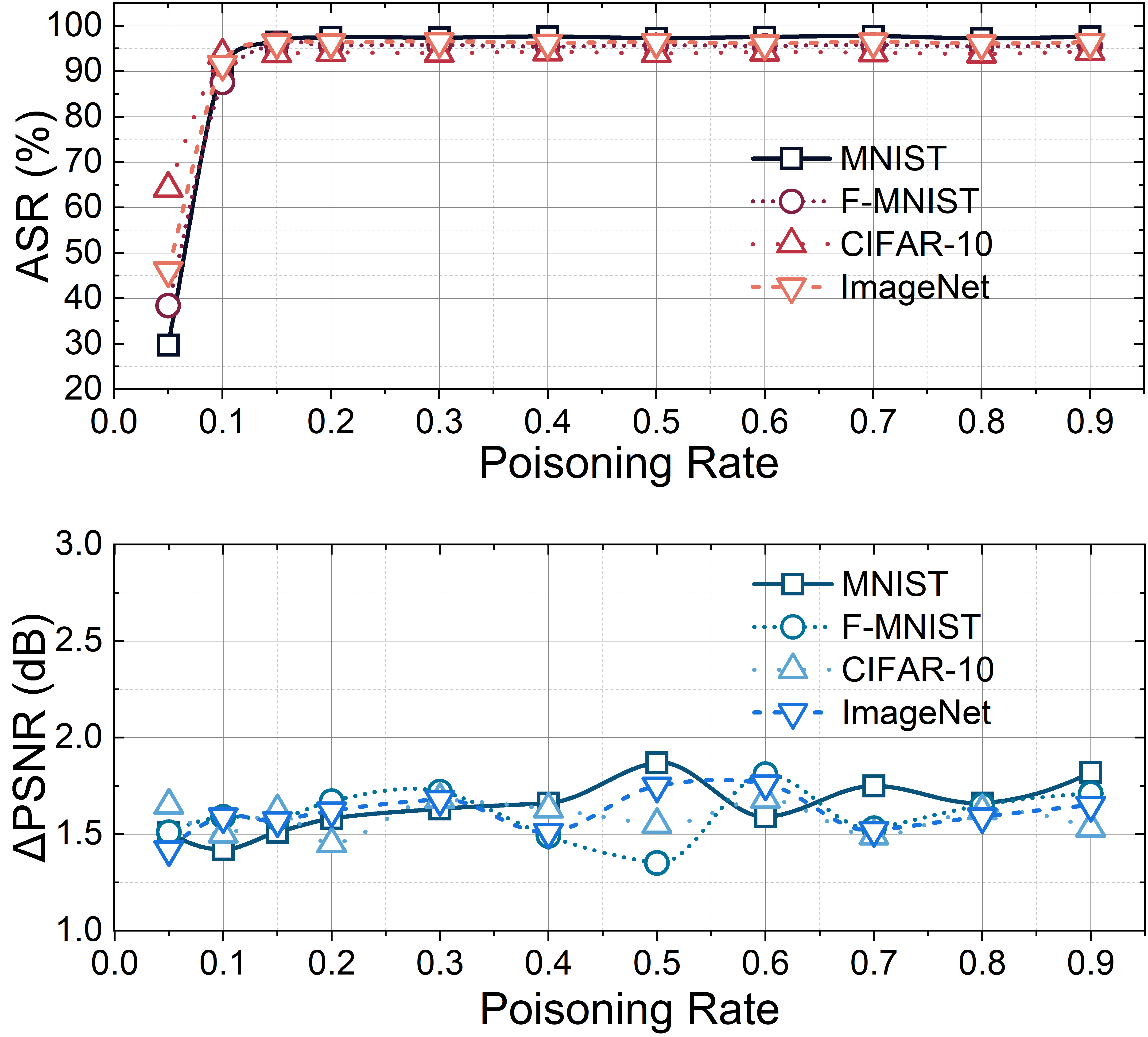}
    \caption{\textbf{Ablation test results across poisoning rates $\gamma$} under communication condition SNR $=25$\,dB  (ASR: $\%$; $\Delta$PSNR: dB). Please \textit{cf.} Sec. \ref{ablation study} for detailed explanations.} 
    \label{fig:poi_rate}
\end{figure}
\begin{figure}[t]
    \centering
    \includegraphics[width=0.95\linewidth, keepaspectratio]{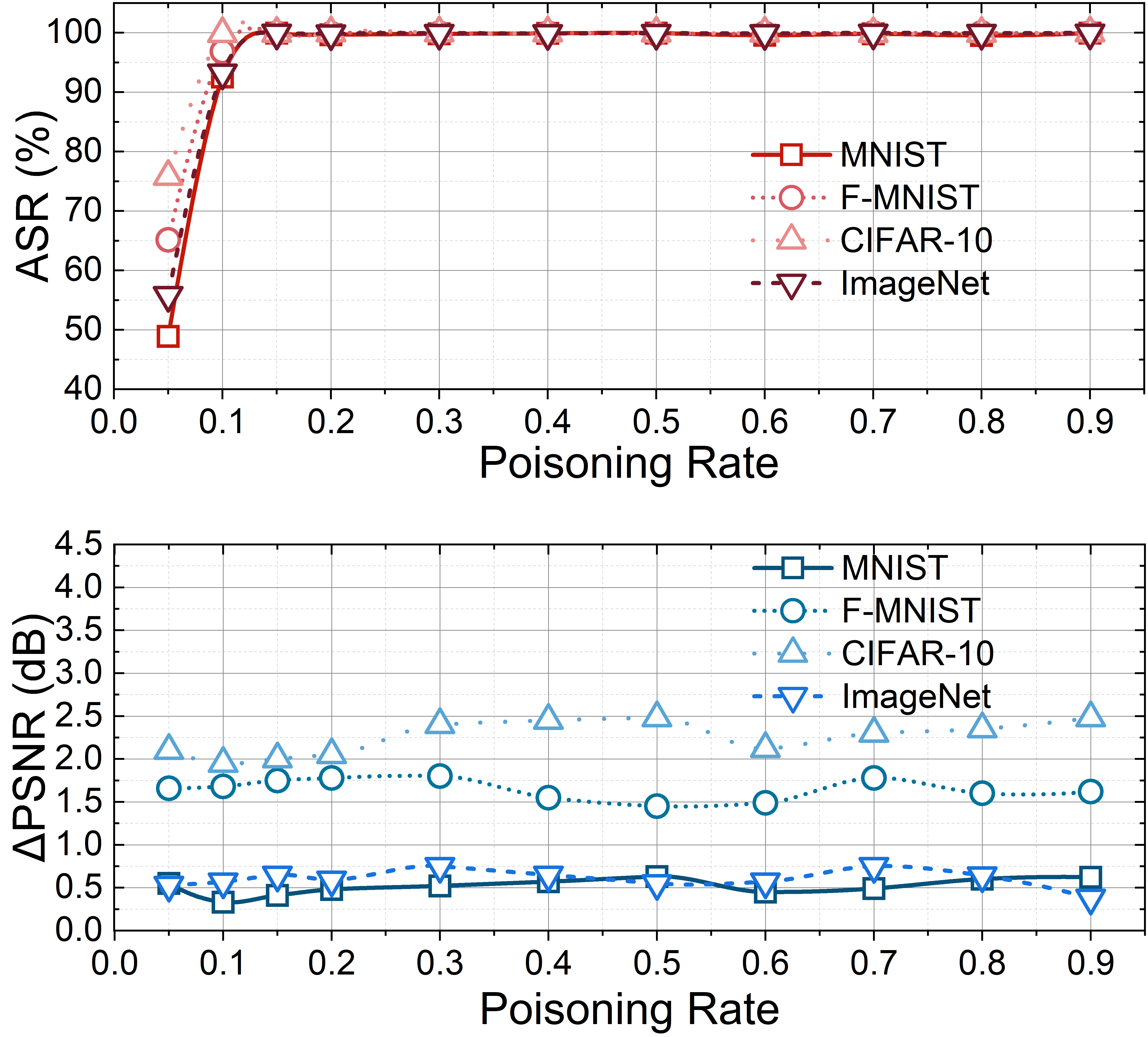}
    \caption{\textbf{Ablation test results across poisoning rates $\gamma$} under communication condition SNR $=5$\,dB  (ASR: $\%$; $\Delta$PSNR: dB). Please \textit{cf.} Sec. \ref{ablation study} for detailed explanations.} 
    \label{fig:poi_rate_5db}
\end{figure}

\vspace{0.8ex}
\noindent \textbf{Poisoning Rate}. 
The rate specifies the fraction of poisoned data during the poisoning-training phase. 
We assessed the minimal levels of poisoning at which our SemBugger keeps effective.
The attack efficacy under varying poisoning rates $\gamma$ was inspected utilizing MNIST, F-MNIST, CIFAR-10, ImageNet datasets under $25$\,dB and $5$\,dB SNR. 
The assessment measures both ASR and transmission fidelity degradation of $\Delta$PSNR. The test results are visualized in Figs. \ref{fig:poi_rate} and \ref{fig:poi_rate_5db}.

In Fig. \ref{fig:poi_rate} of test results under $25$\,dB, $1)$ the success rate increases sharply with the $\gamma$, moving close to $100\%$ for all datasets. Initially, at $\gamma$ $=0.05$, the ASR starts around $50\%$, and then it rises steadily with up-going $\gamma$. 
This finding empirically verifies that our method can deliver considerable ASRs while operating at exceptionally low infection levels.
$2)$ The change in PSNR varies across datasets, but it is still in a minimal range. 
While MNIST retrains low $\Delta$PSNR values, F-MNIST depicts a moderate increase. CIFAR-10 exhibits a more substantial rise in $\Delta$PSNR, especially at higher $\gamma$, whereas ImageNet experiences fluctuating $\Delta$PSNR values. 
Generally, across $\gamma$, our attack strategy preserves victim system's operational integrity with marginal efficiency deviation.

The performance for $4$ datasets uunder $5$\,dB are provided in Fig.~\ref{fig:poi_rate_5db}. 
$1)$ ASR begins at approximately $30\%$ for all datasets at $\gamma = 0.05$ and increases steadily, attaining nearly $97-98\%$ at $\gamma = 0.9$.
The results manifest that our method maintains high effectiveness even under relatively low SNR channel conditions.
$2)$ Regarding $\Delta$PSNR, it conveys slight fluctuations as $\gamma$ increases. 
CIFAR-10 retains the highest $\Delta$PSNR values across most rates, whereas F-MNIST and ImageNet show greater variability, particularly at higher $\gamma$.


\subsection{Defense Study}
\label{defense study}
To further fortify SC system's resilience against SemBugger-type threats, Sec. \ref{defense-framework} introduces a dedicated defensive framework. 
We delved into the defense effectiveness across $5$ SC architectures (\textit{i.e.}, JSCC, JSCC-f, JSCC-q, SCAN, and SemCC) and $4$ benchmark datasets (MNIST, F-MNIST, CIFAR-10, and ImageNet). 
All experiments principally inherited the settings detailed in Sec. \ref{comparison result} and were executed under $25$\,dB and $5$\,dB channel conditions. 
The experimental results are displayed in Tabs. \ref{defense-30db} and \ref{defense-5db}.

Tab. \ref{defense-30db} concludes defense results when operating under $25$\,dB channels. 
Among $5$ SC system architectures evaluated, SemCC drives the ASR down to near $0$\% across all $4$ datasets, while incurring only a $0.20–0.27$\,dB increase in distortion ($\Delta$PSNR). 
The remaining systems also suppress ASR to the low $0.12–0.89$\% range, but residue is still detectable and their $\Delta$PSNR peaks at $0.49$\,dB. 
Wholly, these findings indicate that the proposed defense markedly weakens SemBugger without materially degrading reconstruction quality, with the SemCC configuration acquiring the most pronounced benefit.
The underlying work principle may be that the attack trigger employed in this study is essentially a low-amplitude perturbation. 
By superimposing a carefully crafted, covert noise mask, we can effectively obscure the trigger and distort its statistical signature and realize a robust defense.


\begin{table*}[t]\small
\caption{
\textbf{Defense results against SemBugger backdoors} on MNIST, F-MNIST, CIFAR-10, and ImageNet tasks under standard SNR $= 25$\,dB (ASR$\downarrow$: $\%$; PSNR*$\uparrow$, PSNR$^\circ$, $\Delta$ PSNR $\downarrow$: dB). Please \textit{cf.} Sec. \ref{defense study} for detailed explanations. 
}
\label{defense-30db}
\centering
\scriptsize{
\resizebox{\linewidth}{!}{
\setlength\tabcolsep{3.pt}
\renewcommand\arraystretch{1.1}
\begin{tabular}{r||cccc|cccc|cccc|cccc}
\hline\thickhline
\rowcolor{gray!20}
 & \multicolumn{4}{c|}{\textbf{MNIST Dataset}} & \multicolumn{4}{c|}{\textbf{F-MNIST Dataset}} & \multicolumn{4}{c|}{\textbf{CIFAR-10 Dataset}} & \multicolumn{4}{c}{\textbf{ImageNet Dataset}} \\
\rowcolor{gray!20}
\multirow{-2}{*}{\textbf{Systems}} & ASR & PSNR* & PSNR$^\circ$ & $\Delta$ PSNR & ASR & PSNR* & PSNR$^\circ$ & $\Delta$PSNR & ASR & PSNR* & PSNR$^\circ$ & $\Delta$PSNR & ASR & PSNR* & PSNR$^\circ$ & $\Delta$PSNR \\
\hline\hline
JSCC & 0.12 & 36.19 & 36.47 & 0.28 & 0.00 & 33.63 & 33.82 & 0.19 & 0.23 & 30.94 & 31.15 & 0.21 & 0.00 & 34.71 & 34.93 & 0.22 \\
\hline
\rowcolor{gray!10} JSCC-f & 0.00 & 38.29 & 38.76 & 0.47 & 0.34 & 35.68 & 36.17 & 0.49 & 0.00 & 33.03 & 33.42 & 0.39 & 0.45 & 36.82 & 37.23 & 0.41 \\
\hline
JSCC-q & 0.56 & 38.07 & 38.53 & 0.46 & 0.00 & 35.33 & 35.82 & 0.49 & 0.00 & 32.67 & 33.14 & 0.47 & 0.67 & 36.45 & 36.91 & 0.46 \\
\hline
\rowcolor{gray!10} SCAN & 0.00 & 39.81 & 40.23 & 0.42 & 0.78 & 37.11 & 37.54 & 0.43 & 0.00 & 34.42 & 34.83 & 0.41 & 0.89 & 38.23 & 38.62 & 0.39 \\
\hline
SemCC & 0.00 & 38.93 & 39.20 & 0.27 & 0.00 & 39.07 & 39.34 & 0.27 & 0.00 & 39.58 & 39.82 & 0.24 & 0.00 & 39.42 & 39.62 & 0.20 \\
\hline\thickhline
\end{tabular}}}
\end{table*}

\begin{table*}[t]\small
\caption{
\textbf{Defense results against SemBugger backdoors} on MNIST, F-MNIST, CIFAR-10, and ImageNet tasks under standard SNR $= 5$\,dB (ASR$\downarrow$: $\%$; PSNR*$\uparrow$, PSNR$^\circ$, $\Delta$PSNR$\downarrow$: dB). Please \textit{cf.} Sec. \ref{defense study} for detailed explanations. 
}
\centering
\label{defense-5db}
\scriptsize
\resizebox{\linewidth}{!}{
\setlength\tabcolsep{3pt}
\renewcommand\arraystretch{1.1}
\begin{tabular}{r||cccc|cccc|cccc|cccc}
\hline\thickhline
\rowcolor{gray!20}
 & \multicolumn{4}{c|}{\textbf{MNIST Dataset}}
 & \multicolumn{4}{c|}{\textbf{F-MNIST Dataset}}
 & \multicolumn{4}{c|}{\textbf{CIFAR-10 Dataset}}
 & \multicolumn{4}{c}{\textbf{ImageNet Dataset}}\\
\rowcolor{gray!20}
\multirow{-2}{*}{\textbf{Systems}}
& ASR & PSNR* & PSNR$^\circ$ & $\Delta$PSNR
& ASR & PSNR* & PSNR$^\circ$ & $\Delta$PSNR
& ASR & PSNR* & PSNR$^\circ$ & $\Delta$PSNR
& ASR & PSNR* & PSNR$^\circ$ & $\Delta$PSNR \\
\hline\hline
JSCC & 0.00 & 21.13 & 21.47 & 0.34 & 0.97 & 19.43 & 19.82 & 0.40 & 0.94 & 13.75 & 14.15 & 0.40 & 0.00 & 16.64 & 16.93 & 0.30 \\
\hline
\rowcolor{gray!10} JSCC-f & 0.96 & 23.35 & 23.76 & 0.42 & 0.00 & 20.92 & 21.17 & 0.26 & 0.42 & 16.08 & 16.42 & 0.35 & 0.71 & 18.88 & 19.23 & 0.36 \\
\hline
JSCC-q & 0.53 & 23.12 & 23.53 & 0.41 & 0.82 & 20.66 & 20.82 & 0.17 & 0.00 & 15.97 & 16.14 & 0.18 & 0.29 & 20.55 & 20.91 & 0.36 \\
\hline
\rowcolor{gray!10} SCAN & 0.37 & 22.88 & 23.23 & 0.36 & 0.64 & 20.29 & 20.54 & 0.26 & 0.83 & 15.67 & 15.83 & 0.16 & 0.00 & 19.40 & 19.62 & 0.23 \\
\hline
SemCC & 0.19 & 21.98 & 22.20 & 0.22 & 0.75 & 22.11 & 22.34 & 0.23 & 0.68 & 22.61 & 22.82 & 0.22 & 0.91 & 22.40 & 22.62 & 0.22 \\
\hline\thickhline
\end{tabular}}
\end{table*}

Tab.~\ref{defense-5db} summarizes the defense efficacy under an adverse channel condition of SNR$=5$\,dB.
Across all $5$ SC system architectures and $4$ benchmark datasets, the defense drives the ASR to sub-percent levels ($\sim$ {0\,\%}) and never exceeding {$0.97$\,\%}, wherein it represents a reduction of more than two orders of magnitude relative to the unprotected baseline (\textit{cf.}\ Sec.~\ref{comparison result} for the attack results).
Critically, this security gain is achieved with negligible impact on benign data transmission fidelity.
The post-defense PSNR* differs from the clean-system PSNR$^\circ$ by at most $0.42$\,dB and typically by only $0.2–0.4$\,dB, well below the commonly accepted perceptual threshold.
These results confirm that the defense does not sacrifice communication quality even in low-SNR regimes.

In addition, we verified the certified accuracy of the defense method and summarized it in the \href{https://github.com/youngshallyx/Toward-Polymorphic-Backdoor-against-Semantic-Communication-via-Intensity-Based-Poisoning-Supp-/blob/main/IEEE%20TIFS-Supplementary%20Materials.pdf}{\emph{Supplemental Material}}.

\vspace{0.8ex}
\noindent \textbf{Computation Cost}. 
We show the computational consumption based MNIST and F-MNIST sets. The server is a dual-socket Intel Xeon Platinum 8369B machine (2×32 cores / 64 threads, 128 threads total, 2 NUMA nodes).
It has 10× NVIDIA GeForce GPUs, each with 24GB VRAM (24576 MiB), running Driver 520.61.05 / CUDA 11.8. 
Each sample was processed $10$ times to calculate the average time (w/ SNR $=25$\,dB). We present the average processing time per sample for the entire dataset. The results are listed in Tab. \ref{comp_cost}.

\begin{table}[t]\small
\caption{
\textbf{Average sample running time cost} of defense on MNIST and F-MNIST tasks under standard SNR $= 25$\,dB (ms $\downarrow$). Please \textit{cf.} Sec. \ref{defense study} for detailed explanations. 
}
\centering
\label{comp_cost}
\scriptsize
\resizebox{\linewidth}{!}{
\setlength\tabcolsep{3pt}
\renewcommand\arraystretch{1.1}
\begin{tabular}{r||ccccc}
\hline\thickhline
\rowcolor{gray!20}
 & \multicolumn{5}{c}{\textbf{Test Systems}}\\
\rowcolor{gray!20}
\multirow{-2}{*}{\textbf{Datasets}}
& JSCC & JSCC-f & JSCC-q & SCAN & SemCC \\
\hline\hline
MNIST & 1.15 & 1.22 & 1.93 & 2.07 & 0.94 \\
\hline
\rowcolor{gray!10} F-MNIST & 1.26 & 1.33 & 2.05 & 2.16 & 1.02 \\
\hline\thickhline
\end{tabular}}
\end{table}

As illustrated, the average per-sample processing times lie within a very narrow range. Statistically, the mean processing time across all evaluated systems is only $1.52$\,ms (standard deviation $0.42$\,ms). the maximum latency is merely $2.3\times$ the minimum, and all are far below the inter-frame interval of typical wireless systems (usually $5$--$10$\,ms). The absolute gap between the fastest SemCC architecture and the slowest SCAN architecture is only $1.14$\,ms, which is practically negligible in real deployments. Also, $95\%$ of samples fall within the $0.9$--$2.2$\,ms range, and none of the architectures exceeds the $3$\,ms latency budget commonly tolerated by real-time systems. This is because the proposed defense only requires injecting smoothed Gaussian noise and does not introduce additional system processing. Hence, although architectural differences lead to slight variations in runtime, these differences are minor and well within the tolerance of practical communication systems, with no obstacle to large-scale deployment.

\vspace{0.8ex}
\noindent \textbf{Downstream Task Influence}. 
To further assess the task-level impact of the proposed defense in practical deployments, we cascade a downstream classifier (RegNetY-16GF network) with the receiver-side output of the SC architecture to form an end-to-end transmission–identification system, on which we implement both attacks and defenses. Specifically, we measure the system’s baseline classification performance without attacks, as well as its efficacy after using the defense, and quantify the task-level impact of our defense.
The experiments were implemented on MNIST dataset under SNR $=25$\,dB. Tab. \ref{def_task_influence} summarizes the evaluation results.

\begin{table}[t]\small
\caption{
\textbf{Task classification efficacy} of defense on MNIST tasks under standard SNR $= 25$\,dB ($\%$ $\uparrow$). Please \textit{cf.} Sec. \ref{defense study} for detailed explanations. 
}
\centering
\label{def_task_influence}
\scriptsize
\resizebox{\linewidth}{!}{
\setlength\tabcolsep{3pt}
\renewcommand\arraystretch{1.1}
\begin{tabular}{r||ccccc}
\hline\thickhline
\rowcolor{gray!20}
 & \multicolumn{5}{c}{\textbf{Test Systems}}\\
\rowcolor{gray!20}
\multirow{-2}{*}{\textbf{Condition}}
& JSCC & JSCC-f & JSCC-q & SCAN & SemCC \\
\hline\hline
Clean & 95.15 & 95.22 & 94.93 & 94.07 & 95.94 \\
\hline
\rowcolor{gray!10} Defense & 94.96 & 95.03 & 94.71 & 93.92 & 95.82 \\
\hline\thickhline
\end{tabular}}
\end{table}

The experimental results manifest minimal performance degradation introduced by the defense scheme. The classification accuracy of SC shows an average reduction of only $0.17$ (from $95.06\%$ to $94.89\%$), with the maximum reduction being $0.22$ percentage points (JSCC-q system: $94.93\% \rightarrow 94.71\%$) and the minimum reduction being $0.12$ percentage points (SemCC system: $95.94\% \rightarrow 95.82\%$). All systems maintain accuracy above $93.92\%$ after defense implementation, with relative differences less than $0.3\%$ compared to the clean counterparts. This confirms that the defense preserves over $98\%$ of semantic information integrity while introducing only marginal PSNR degradation during transmission, with downstream classification tasks virtually unaffected. 


\section{Conclusion}
In this paper, we first identify the key limitation in current research on SC backdoors: the prevailing monomorphic activation paradigm hampers efficiency, adaptability, and output variety. 
To address these constraints, \textit{we introduce the first polymorphic backdoor in the SC field, which dynamically modulates the intensity of implicit triggers embedded in the transmission inputs, hence allowing adversaries to manipulate the system to yield multiple, distinct reconstruction targets}.
Specifically, the polymorphic backdoor is implanted into the shared knowledge of the SC system through a multi-effect poisoning framework. This preserves the fidelity of benign transmissions while enabling the system to distinguish graded covert triggers and yield adversarial reconstruction outputs uniquely mapped to each trigger intensity.
Additionally, to counter the proposed backdoor, we devise a provably secure semantic-smoothing defense. Based on a formally derived lower defense bound, the scheme injects carefully calibrated noise into inputs, whereby it attenuates latent trigger signals and consequently prevents backdoor activations.
Extensive experiments demonstrate that our proposed attack method achieves high attack efficacy (ASR$ > 90\%$) while introducing minimal degradation to the system's normal functionality ($\Delta\text{PSNR} < 2\,\text{dB}$). Furthermore, our designed defense scheme can effectively mitigate the attack, reducing the success rate to a negligible level ($\text{ASR} < 1\%$).

Our future research will generalize the current framework to encompass distributed poisoning attacks, particularly examining cross-client telecom patterns and their countermeasures.

\section{Acknowledgment}
This work is funded by the National Natural Science Foundation of China (Nos. 62572314, 62471301 and U21B2019). 
Gaolei Li and Jun Wu are the corresponding authors.

\bibliographystyle{ieeetr}
\bibliography{ref}

\vfill

\end{document}